\documentclass[structabstract]{aa}  
\usepackage{graphicx}
\usepackage{txfonts}
\begin{document}
\title{The seismic properties of low-mass He-core white dwarf stars}
 
\subtitle{}

\author{A. H. C\'orsico\inst{1,2},  
        A. D. Romero\inst{1,2}, 
        L. G. Althaus\inst{1,2} \and 
        J. J. Hermes\inst{3,4}}
\institute{$^{1}$ Facultad de Ciencias Astron\'omicas y Geof\'{\i}sicas, 
           Universidad Nacional de La Plata, Paseo del Bosque s/n, 1900 
           La Plata, Argentina\\
           $^{2}$ CCT La Plata, CONICET, 1900 La Plata, Argentina\\
           $^{3}$ Department of Astronomy, University of Texas at Austin,
           Austin, TX- 78712, USA\\
           $^{4}$  McDonald Observatory, Fort Davis, TX- 79734, USA\\
           \email{acorsico,aromero,althaus@fcaglp.unlp.edu.ar; jjhermes@astro.as.utexas.edu}     
           }
\date{Received ; accepted }

\abstract {In recent years,  many low-mass ($\lesssim 0.45 M_{\odot}$)
  white dwarf stars expected to harbor  He cores have been detected in
  the field of the Milky Way and in several galactic globular and open
  clusters.   Until   recently,  no   objects  of  this   kind  showed
  pulsations.  This  situation has changed recently  with the exciting
  discovery of SDSS  J184037.78+642312.3, the first pulsating low-mass
  white dwarf star.}  {Motivated  by this extremely important finding,
  and in  view of the  very valuable asteroseismological  potential of
  these objects, we present  here a detailed pulsational study applied
  to low-mass He-core white dwarfs  with masses ranging from $0.17$ to
  $0.46 M_{\odot}$,  based on full  evolutionary models representative
  of  these objects.   This study  is aimed  to provide  a theoretical
  basis  from  which  to  interpret future  observations  of  variable
  low-mass white dwarfs.}  {The background stellar models on which our
  pulsational  analysis was carried  out were  derived by  taking into
  account the  complete evolutionary history of  the progenitor stars,
  with special  emphasis on the diffusion processes  acting during the
  white dwarf cooling phase. We computed nonradial $g$-modes to assess
  the dependence  of the pulsational properties of  these objects with
  stellar  parameters  such as  the  stellar  mass  and the  effective
  temperature,  and also  with element  diffusion processes.   We also
  performed a $g$- and  $p$-mode pulsational stability analysis on our
  models and found well-defined  blue edges of the instability domain,
  where these  stars should start  to exhibit pulsations.}   {We found
  substantial differences  in the  seismic properties of  white dwarfs
  with $M_*  \gtrsim 0.20 M_{\odot}$ and the  extremely low-mass (ELM)
  white dwarfs ($M_* \lesssim 0.20 M_{\odot}$).  Specifically, $g$-mode
  pulsation modes  in ELM white  dwarfs mainly probe the  core regions
  and are  not dramatically affected  by mode-trapping effects  by the
  He/H  interface,  whereas the  opposite  is  true  for more  massive
  He-core  white dwarfs.   We found  that element  diffusion processes
  substantially  affects the  shape  of the  He/H chemical  transition
  region, leading to non-negligible  changes in the period spectrum of
  low-mass white dwarfs, in particular  in the range of stellar masses
  characteristic  of  ELM  objects.  Finally, our  stability  analysis
  successfully  predicts the  pulsations  of the  only known  variable
  low-mass  white  dwarf   (SDSS  J184037.78+642312.3)  at  the  right
  effective  temperature,  stellar  mass  and range  of  periods.}{Our
  computations   predict   both    $g$-   and   $p$-mode   pulsational
  instabilities  in a  significant number  of known  low-mass  and ELM
  white  dwarfs.   It is  worth  observing  these  stars in  order  to
  discover if they pulsate.}

\keywords{asteroseismology -- stars: oscillations -- stars: white dwarfs 
-- stars: evolution -- stars: interiors}
\titlerunning{Pulsating low-mass white dwarfs}
   \maketitle
%

\section{Introduction}
\label{introduction}

White dwarf  stars are the most  common evolutionary fate  of low- and
intermediate-mass  ($M_* \lesssim  11 M_{\odot}$;  Siess  2007) stars.
Indeed, the  vast majority ---more than  $95 \%$--- of  all stars will
die as white dwarfs.  As  such, these old and compact stellar remnants
provide a  wealth of  information about the  evolution of  stars, star
formation, and  the age of a  variety of stellar  populations, such as
our  Galaxy and  open and  globular clusters  (Winget \&  Kepler 2008;
Garc\'ia-Berro et al. 2010; Althaus et al. 2010).  Among the different
flavors of white  dwarfs, the most common is the  spectral class of DA
white dwarfs, characterized by H rich envelopes, that comprises around
$80 \%$ of all known white dwarfs.  An important property of the white
dwarf population is their mass distribution.  For DA white dwarfs, the
mass distribution  peaks at $\approx  0.59 M_{\odot}$, and  exhibits also
high-mass and low-mass components (Kepler et al. 2007; Tremblay et al.
2011; Kleinman et al.  2012).  The population of low-mass white dwarfs
has  masses  lower than  $0.45  M_{\odot}$  and  peaks at  $\approx  0.39
M_{\odot}$.  Recently,  a large number  of low-mass white  dwarfs with
masses  below  $0.20-0.25 M_{\odot}$  has  been  discovered (Kawka  \&
Vennes 2009; Brown et al.  2010, 2012; Kilic et al.  2011, 2012); they
are referred to as extremely low-mass (ELM) white dwarfs.

The  low-mass white dwarf  population is  probably produced  by strong
mass-loss  episodes at  the red  giant branch  (RGB) phase  before the
He-flash onset. As such, these  white dwarfs are expected to harbor He
cores,  in contrast  to average  mass white  dwarfs, which  all likely
contain  C/O  cores.  For   solar  metallicity  progenitors  ($Z  \approx
0.01-0.02$), mass-loss  episodes must occur in  binary systems through
mass-transfer, since single star evolution  is not able to predict the
formation of these stars in a Hubble time.  This evolutionary scenario
is confirmed by the fact that  most of low-mass white dwarfs are found
in  binary  systems  (e.g.,  Marsh  et  al.   1995),  and  usually  as
companions  to millisecond pulsars  (van Kerkwijk  et al.   2005).  In
particular, binary evolution  is the most likely origin  for ELM white
dwarfs (Marsh  et al. 1995).   On the other hand,  for 
high-metallicity progenitors ($Z  \approx 0.03-0.05$), the  He-core 
flash can  probably be
avoided by mass  losses due to strong stellar winds  on the RGB.  This
can be the origin of  the isolated low-mass He-core white dwarfs ($M_*
\approx 0.4 M_{\odot}$)  found in the high-metallicity open  cluster NGC 6791
(Hansen 2005; Kalirai et al. 2007; Garc\'ia-Berro et al. 2010).

The evolution of low-mass white  dwarfs is strongly dependent on their
stellar  mass and the  occurrence of  element diffusion processes.   
Althaus et
al. (2001)  and Panei  et al.  (2007)  have demonstrated  that element
diffusion  leads to  a  dichotomy  regarding the  thickness  of the  H
envelope, which  translates into  a dichotomy in  the age  of low-mass
He-core  white  dwarfs.  Specifically,  for  stars  with $M_*  \gtrsim
0.18-0.20 M_{\odot}$, the  white dwarf progenitor experiences multiple
diffusion-induced  thermonuclear  flashes  that  burn most  of  the  H
content of the envelope, and as a result, the remnant enters the final
cooling track  with a very thin H  envelope. In this way,  the star is
unable  to  sustain  stable   nuclear  burning  while  cools  and  the
evolutionary    timescale    is    rather   short    ($\approx   10^{7}$
yr)\footnote{Note  that,  if  element diffusion processes are not included,
  theoretical  computations  also predict  the  occurrence of  H-shell
  flashes before  the terminal  cooling branch is  reached, but  the H
  envelopes remain  thick, with substantial H burning  and large white
  dwarf cooling ages  (Driebe et al.  1998; Sarna  et al. 2000).}.  On
the  other hand,  if  $M_* \lesssim  0.20  M_{\odot}$, the  white dwarf
progenitor  does not  experience H  flashes  at all,  and the  remnant
enters its  terminal cooling branch with  a thick H  envelope. This is
thick enough for residual H  nuclear burning to become the main energy
source, that ultimately slows down the evolution, in which case the cooling
timescale is  of the order $\approx 10^{9}$ yrs.  The  age dichotomy has
been also  suggested by observations  of those low-mass  He-core white
dwarfs that are companions to  millisecond pulsars (Bassa et al. 2003;
Bassa 2006).

The internal structure of white dwarfs can be disentangled by means of
asteroseismology (Winget  \& Kepler  2008; Fontaine \&  Brassard 2008;
Althaus et al.  2010). White dwarf asteroseismology allows us to place
constraints on  the stellar mass,  the thickness of  the compositional
layers,  and  the  core  chemical composition,  among  other  relevant
properties.  In connection with  the core composition of white dwarfs,
some theoretical  work in the  past has explored the  differences that
should  be expected  in  the pulsation  properties  of low-mass  white
dwarfs harboring cores made either of C/O or He (Althaus et al. 2004),
and  high mass  white dwarfs  with cores  made either  of C/O  or O/Ne
(C\'orsico et al. 2004).   More recently, Castanheira \& Kepler (2008,
2009),  and   Romero  et  al.    (2012)  (see also Romero 2012)  
have  carried   out  detailed
asteroseismological studies  on a large  number of pulsating  DA white
dwarfs  (DAV or  ZZ  Ceti variables),  providing valuable  information
about the range  of the H envelope thicknesses  expected for the class
of DA white dwarfs.

In order to  apply the principles of asteroseismology  on low-mass and
ELM white  dwarfs, it is necessary  to find such stars  (either in the
Galactic  field   or  in  stellar   clusters)  undergoing  pulsations,
something that  until very  recently had not  been possible.   In this
regard,  it is  worth  mentioning that  exhaustive  searches reported  by
Steinfadt  et  al.  (2012)  have  given  null  results, despite  their
theoretical  predictions (Steinfadt et  al.  2010,  hereinafter SEA10)
suggesting  pulsational  instabilities at  least  in  stars with  $M_*
\lesssim  0.20  M_{\odot}$  (ELM  white  dwarfs).   The  situation  has
improved   drastically   with   the   exciting   discovery   of   SDSS
J184037.78$+$642312.3, the first pulsating  ELM white dwarf (Hermes et
al. 2012).   This star ($T_{\rm eff}=  9100 \pm 170$ K,  $\log g= 6.22
\pm  0.06$)  exhibits  multiperiodic  photometric  variations  with  a
dominant  period  at $\approx  4698$  s,  much  longer than  the  longest
periodicities  detected  in any  ZZ  Ceti  star  ($100-1200$ s).   The
discovery of this  pulsating ELM white dwarf opens  the possibility of
sounding the interiors of low-mass white dwarfs employing the tools of
asteroseismology.

Needless to say, accurate  and realistic stellar models representative
of low-mass white dwarfs are needed to correctly interpret the present
and  future   observations  of   pulsations  in  these   objects.   In
particular, the complete evolutionary  history of the progenitor stars
must  be taken  fully  into account  in  order to  assess the  correct
thermo-mechanical structure  of the white  dwarf and the  thickness of
the H envelope remaining  after multiple diffusion-induced CNO flashes
occurring  before  the  terminal  cooling  branch  is  reached,  which
strongly determines the  cooling timescales.  Also, 
element diffusion processes must be considered  in order to consistently 
account  for the evolving
shape  of the  internal  chemical profiles  (and,  in particular,  the
chemical   transition   regions)   during  the   white dwarf   cooling
phase. Finally,  stable H burning,  which is particularly  relevant in
the case  of ELM white dwarfs (and  can play an important  role in the
excitation of pulsations), must be taken into account as well.

Motivated by  the asteroseismic potential of  pulsating low-mass white
dwarfs,  and stimulated  by the  discovery of  the first  variable ELM
white dwarf, we have started in the La Plata Observatory a theoretical
study of  the pulsation properties  of low-mass, He-core  white dwarfs
with masses in the range  $0.17-0.46 M_{\odot}$. Our interest in these
stars  is  strengthened by  the  availability  of  detailed and  fully
evolutionary  stellar  models  extracted  from  the  sequences  of
low-mass    He-core   white   dwarfs    presented   by    Althaus   et
al. (2009). These models  were derived by considering the evolutionary
history of progenitor stars  with high metallicity, and taking
into  account  a  self-consistent,  time-dependent  treatment  of  the
gravitational settling and chemical  diffusion, as well as of residual
nuclear burning.   In this study,  we explore the  adiabatic pulsation
properties of these models, that is, the expected range of periods and
period  spacings,  the propagation  characteristics  of the  pulsation
modes, the regions  of period formation, as well  as the dependence on
the effective  temperature and stellar  mass.  For the first  time, we
assess the pulsation properties of He-core white dwarfs with masses in
the  range $0.20-0.45  M_{\odot}$.   In particular,  we highlight  the
expected differences  in the seismic  properties of objects  with $M_*
\gtrsim 0.20 M_{\odot}$  and the ELM white dwarfs  ($M_* \lesssim 0.20
M_{\odot}$).  Since  the most  important  work  in  the literature  on
pulsations of  low-mass white dwarfs is  the study of  SEA10, we shall
invoke it repeatedly  to compare their results with  our own findings,
in particular  regarding the seismic  properties of ELM  white dwarfs.
We also explore  the role of time-dependent element  diffusion in ELM
models  on their pulsational  properties, an  aspect not  discussed by
SEA10. In addition,  we compute $g$- and $p$-mode blue edges  
of the instability domain of  these  stars through  a  nonadiabatic  
stability  analysis of  our
sequences,  and  in  particular,  we  try  to  determine  whether  our
computations  are able  to predict  the pulsations  exhibited  by SDSS
J184037.78$+$642312.3.  This paper is  organized as follows.  In Sect.
\ref{computational_tools} we briefly  describe our numerical tools and
the main  ingredients of the  evolutionary sequences we use  to assess
the pulsation  properties of low-mass He-core white  dwarfs.  In Sect.
\ref{pulsation_results}  we present in  detail our  adiabatic
pulsation results, while in Sect. \ref{sdss} we describe 
our nonadiabatic results and the case of SDSS J184037.78$+$642312.3.
Finally, in Sect. \ref{conclusions}  we summarize the main findings of
the paper.

\begin{figure*}
\centering
\includegraphics[clip, width=0.95\textwidth, angle=0]{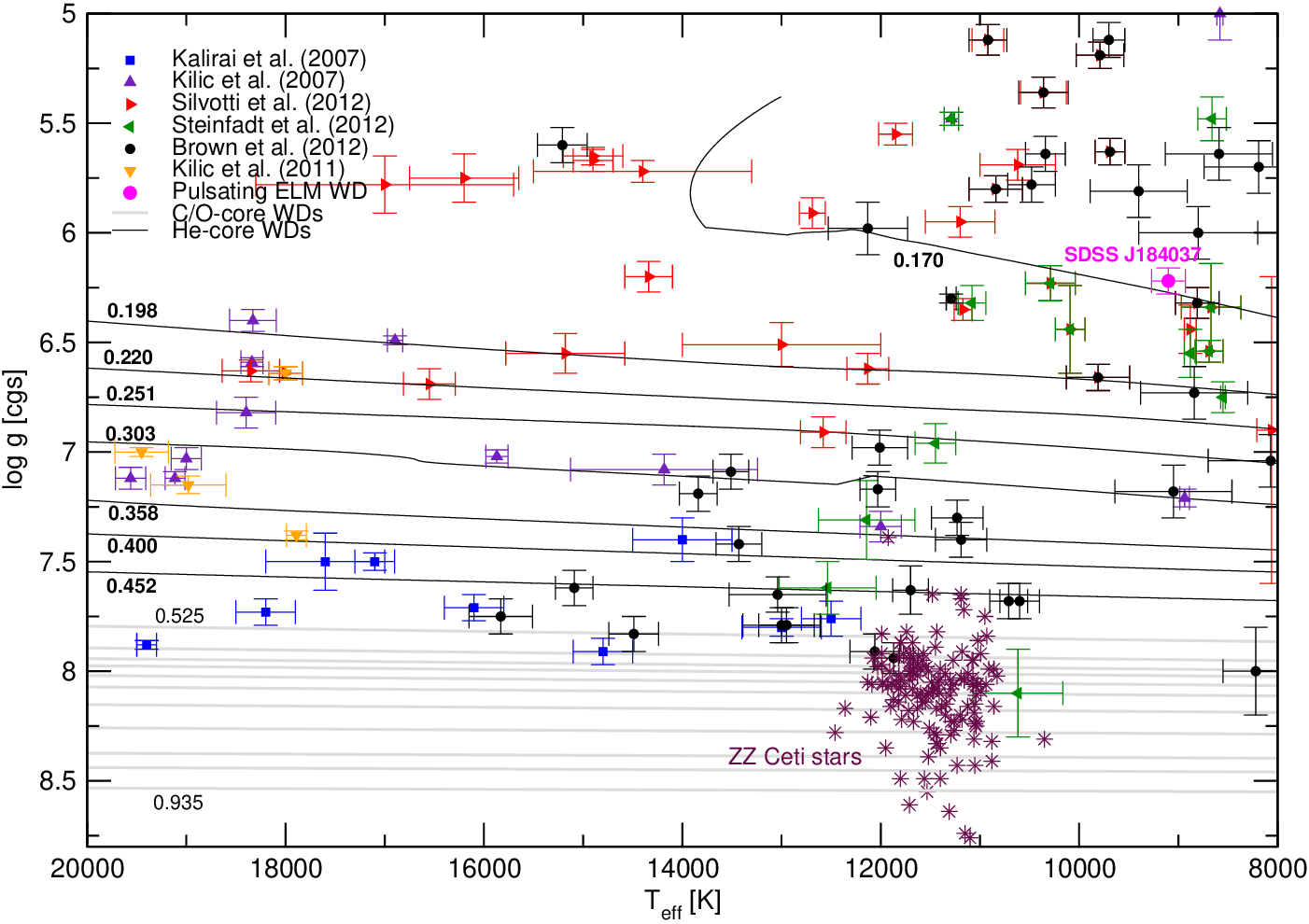}
\caption{The  $T_{\rm eff}  -  \log g$  diagram  showing our  low-mass
  He-core  white dwarf evolutionary  tracks (black  thin  lines). Bold
  numbers  correspond to  the  stellar mass  of  each sequence.   The
  different  symbols with  error bars  represent the  location  of the
  low-mass white  dwarfs known  to date. We  include field  objects as
  well as stars  found in the open cluster NGC 6791  to have a general
  overview of  the observational status of these  stars.  The location
  of the first known pulsating ELM white dwarf ($T_{\rm eff}= 9100 \pm
  170$ K, $\log  g= 6.22 \pm 0.06$; Hermes et  al.  2012) is displayed
  with a  small circle  (magenta).  As reference,  we also  depict the
  C/O-core  white dwarf  evolutionary tracks  (gray thick lines)  for
  masses between  $0.525$ and $0.935  M_{\odot}$ (Romero et  al.  2012)
  and the location of the known ZZ Ceti (DAV) stars.}
\label{tracks}
\end{figure*}

\section{Computational tools and evolutionary sequences}
\label{computational_tools}

\subsection{Evolutionary code and input physics}

The  evolutionary models  employed  in our  pulsational analysis  were
generated with  the LPCODE evolutionary code,  which produces complete
and detailed  white dwarfs models incorporating  very updated physical
ingredients.  While detailed information  about LPCODE can be found in
Althaus et al. (2005, 2009)  and references therein, we highlight here
only  those  ingredients  which  are  important for  our  analysis  of
low-mass, He-core white dwarf stars:

\begin{itemize}

\item  [(a)] The standard  mixing length  theory (MLT)  for convection
  (with the free parameter $\alpha  = 1.6$) has been adopted. Had
  we  adopted a  different convective  efficiency, then  the adiabatic
  pulsation     results    presented     in    this     work    (Sect.
  \ref{pulsation_results})  should  not   be  affected,  although  the
  predicted blue  edge of  the ELM and  low-mass He-core  white dwarfs
  instability domain (Sect.  \ref{sdss}) should appreciably change.
 
\item [(b)] A supersolar metallicity for the progenitor stars has been
  considered:  $Z  =  0.03$.  

\item [(c)] Radiative opacities for arbitrary metallicity in the range
  from 0  to 0.1 are from  the OPAL project (Iglesias  \& Rogers 1996)
  supplemented  at low  temperatures with  the molecular  opacities of
  Alexander \& Ferguson (1994).

\item   [(d)]   Neutrino  emission   rates   for   pair,  photo,   and
  bremsstrahlung processes were taken from Itoh et al. (1996), and for
  plasma processes we included the treatment of Haft et al. (1994).

\item [(e)] Conductive opacities are those of Cassisi et al. (2007).

\item [(f)] For the white dwarf  regime we employed an updated version
  of the Magni \& Mazzitelli (1979) equation of state.

\item [(g)] The nuclear network  takes into account 16 elements and 34
  nuclear reactions  for pp  chains, CNO bi-cycle,  He burning,  and C
  ignition.

\item [(h)] Time-dependent diffusion due to gravitational settling and
  chemical  and thermal diffusion  of nuclear  species was  taken into
  account  following  the  multicomponent  gas  treatment  of  Burgers
  (1969).

\item  [(i)]  Abundance changes  were  computed  according to  element
  diffusion, nuclear reactions,  and convective mixing.  This detailed
  treatment  of abundance  changes by  different processes  during the
  white dwarf regime constitutes a key aspect in the evaluation of the
  importance of  residual nuclear burning for the  cooling of low-mass
  white dwarfs.

\end{itemize}

\subsection{Pulsation codes}

In this work, we have carried out a detailed adiabatic $g$-mode pulsation study
aimed  at  exploring  the   seismic  properties  of  low-mass  He-core
white dwarf  models.  In  addition,  we  have  performed  a  $g$-
and $p$-mode 
stability
analysis of  our model sequences.   The pulsation computations  of our
adiabatic survey  were performed with the pulsation  code described in
detail in C\'orsico \& Althaus  (2006), which is coupled to the LPCODE
evolutionary  code.   The  pulsation   code  is  based  on  a  general
Newton-Raphson  technique that  solves  the full  fourth-order set  of
equations  and   boundary  conditions  governing   linear,  adiabatic,
nonradial stellar  pulsations following the  dimensionless formulation
of Dziembowski  (1971) (see Unno  et al.  1989).  The  prescription we
follow to  assess the run  of the Brunt-V\"ais\"al\"a  frequency ($N$)
for a degenerate  environment typical of the deep  interior of a white
dwarf is  the so-called ``Ledoux Modified'' treatment  (Tassoul et al.
1990).

As for the stability  analysis, we have employed the finite-difference
nonadiabatic pulsation  code described in  detail in C\'orsico  et al.
(2006).   The  code solves  the  full  sixth-order  complex system  of
linearized equations and  boundary conditions as given by  Unno et al.
(1989).  Typically, our adiabatic and non-adiabatic periods differ
  by less than  $\approx 2 \%$.  The caveat of our  analysis is that the
nonadiabatic computations rely on the frozen-convection approximation,
in which the perturbation of  the convective flux is neglected.  While
this  approximation is  known  to give  unrealistic  locations of  the
$g$-mode red edge of instability, it leads to satisfactory predictions
for the  location of the  blue edge of  the ZZ Ceti  (DAV) instability
strip  (see,  e.g., Brassard  \&  Fontaine  1999  and van  Grootel  et
al. 2012) and also for the  V777 Her (DBV) instability strip (see, for
instance, Beauchamp et al.  1999 and C\'orsico et al. 2009).

\subsection{Evolutionary sequences}

To derive  starting configurations  for the He-core  cooling sequences
consistent  with  the evolutionary  history  of  the progenitor  star,
Althaus et al.  (2009) simply  removed mass from a $1 M_{\odot}$ model
at  the appropriate  stages of  its evolution  (Iben \&  Tutukov 1986;
Driebe et al.  1998). Other  details about the procedure to obtain the
initial models are  provided in Althaus et al.   (2009). The resulting
final   stellar   masses  ($M_*/M_{\odot}$)   are   listed  in   Table
\ref{tabla1}  with the  total amount  of H  contained in  the envelope
($M_{\rm   H}/M_*$),  and   the  coordinate   $-\log(q)$   (being  $q=
1-M_r/M_*$)  of the  location of  the He/H  chemical  interface (where
$X_{\rm  He}= X_{\rm  H}= 0.5$),  for models  at $T_{\rm  eff} \approx
10\,000$ K. The last column shows  the time spent by the stars to cool
from $T_{\rm eff} \approx 14\, 000$ K to $\approx 8000$ K.

We mention  again that a metallicity  of $Z= 0.03$  for the progenitor
stars  has been adopted.  This value  is appropriate  for overmetallic
environments such as the open  cluster NGC 6791 (Althaus et al. 2009).
Had we adopted a solar metallicity value ($Z \approx 0.01-0.02$) instead,
then all our  sequences would be characterized by  H envelopes thicker
than those considered here  (second column of Table \ref{tabla1}), but
it would not qualitatively affect the results presented in this paper.

The
sequence of  $M_*= 0.17 M_{\odot}$ does  not correspond to  the set of
sequences  presented by  Althaus et  al. (2009);  it  was specifically
computed  for the  present  study.  The models  of  this sequence  are
representative of ELM white  dwarfs.  The evolutionary history of this
sequence is quite different  as compared with the remaining sequences,
because its progenitor  star does not experience any  CNO flash.  As a
result,  the remaining  H envelope  is markedly  thicker than  for the
other  sequences.  The very  distinct origin  of the  $0.17 M_{\odot}$
sequence is also emphatically evidenced by the time the star spends to
cool from $\approx 14\,000$ to  $\approx 8000$ K, which is about $3.7$
times longer than for the case of the $0.198 M_{\odot}$ sequence.

\begin{table}
\centering
\caption{Selected  properties  of our  He-core  white dwarf  sequences
  (metallicity  of the  progenitor star:  $Z= 0.03$)  at  $T_{\rm eff}
  \approx 10\,000$  K: the stellar  mass, the mass  of H in  the outer
  envelope, the location of the  He/H chemical interface, and the time
  it takes the  star to cool from $T_{\rm eff} \approx  14\, 000$ K to
  $\approx 8000$ K.}
\begin{tabular}{cccc}
\hline
\hline
\noalign{\smallskip}
 $M_*/M_{\odot}$  & $M_{\rm H}/M_*\ [10^{-3}]$ & $-\log(1-M_r/M_*)$ & 
$\tau\ [{\rm Myr}= 10^6 {\rm yr}]$ \\
\hline
\noalign{\smallskip}
 0.170 & 12.992 &  2.0196 & 2122.394\\ 
 0.198 & 3.7717 &  3.2851 & 576.575\\
 0.220 & 1.6621 &  3.8237 & 387.775\\
 0.251 & 1.2039 &  3.1733 & 580.847\\
 0.303 & 1.2661 &  3.0191 & 798.432\\
 0.358 & 0.5428 &  3.3759 & 917.173\\
 0.400 & 0.6638 &  3.2590 & 1204.545\\
 0.452 & 0.3708 &  3.5459 & 1228.023\\
\noalign{\smallskip}
\hline
\hline
\end{tabular}
\label{tabla1}
\end{table}

\begin{figure}
\centering
\includegraphics[clip, width=0.45\textwidth, angle=0]{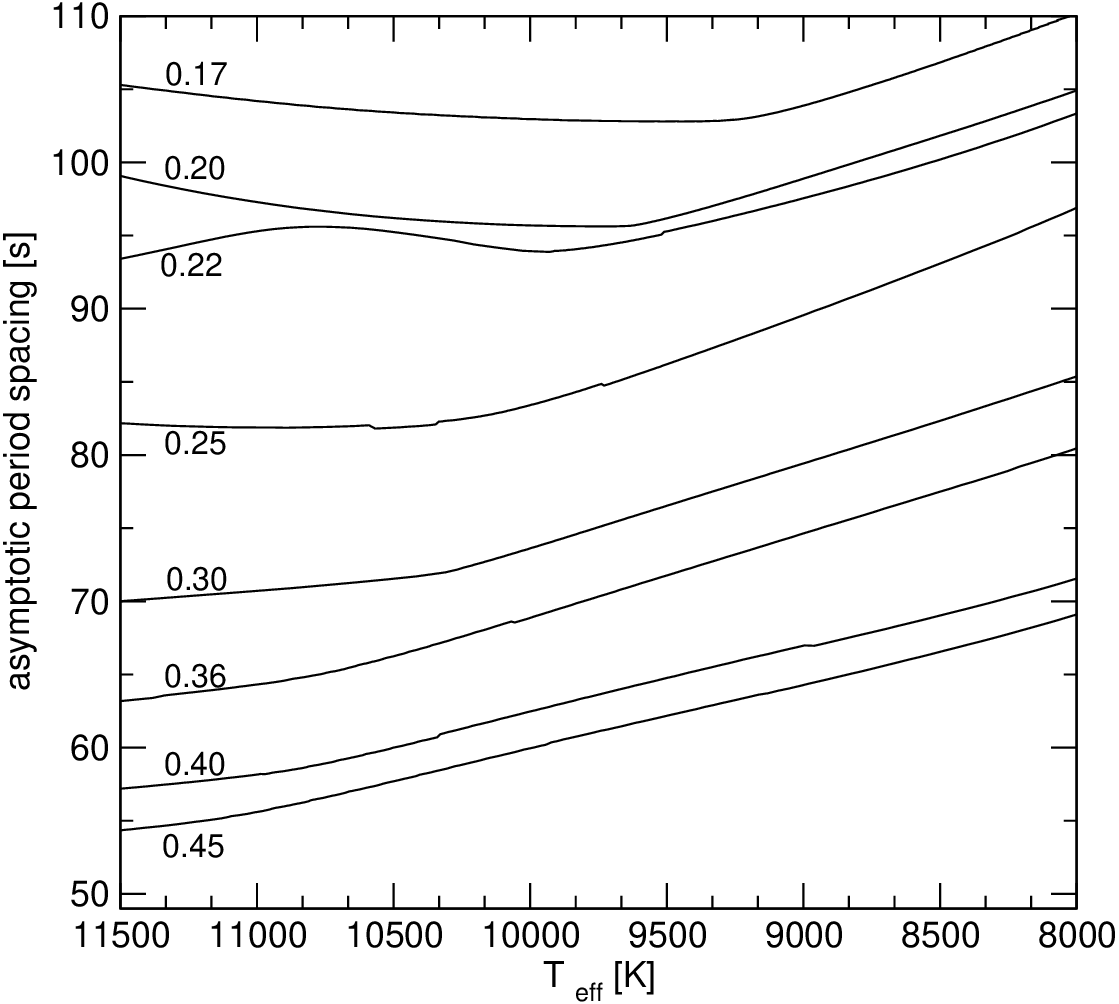}
\caption{The  asymptotic  period spacing  in  terms  of the  effective
  temperature for all of our low-mass He-core evolutionary sequences.}
\label{asint}
\end{figure}

The long  time that ELM white  dwarfs should spend  at these effective
temperatures, by virtue of vigorous stable H burning via the pp chain,
motivated  Steinfadt et al.   (2012) to  search for  pulsating objects
with masses  lower than $\approx  0.20 M_{\odot}$. We note  from Table
\ref{tabla1}, however, that  models with $\approx 0.40-0.45 M_{\odot}$
should  also have  comparable evolutionary  timescales.   This implies
that there should  be a good chance of  finding pulsating objects with
these masses, if they exist, and current searches of variable low-mass
white dwarfs should not be restricted only to ELM white dwarfs.

In Fig.  \ref{tracks}  we depict our evolutionary tracks  on the plane
$T_{\rm  eff} -  \log  g$ (black  thin  lines). In  order  to have  an
overview of the observational status, we have included the location of
the low-mass white dwarfs known to date. They include field stars that
are members of  binary systems (Kilic et al. 2007,  2011; Brown et al.
2012; Steinfadt  et al. 2012, Silvotti  et al. 2012)  and single stars
residing  in the  open cluster  NGC 6791  (Kalirai et  al.  2007).  As
reference, we include also the location of the known ZZ Ceti stars and 
the evolutionary tracks corresponding to C/O-core white dwarfs 
(gray thick lines). 

\section{Adiabatic pulsation properties} 
\label{pulsation_results}

\begin{figure}
\centering
\includegraphics[clip, width=0.45\textwidth, angle=0]{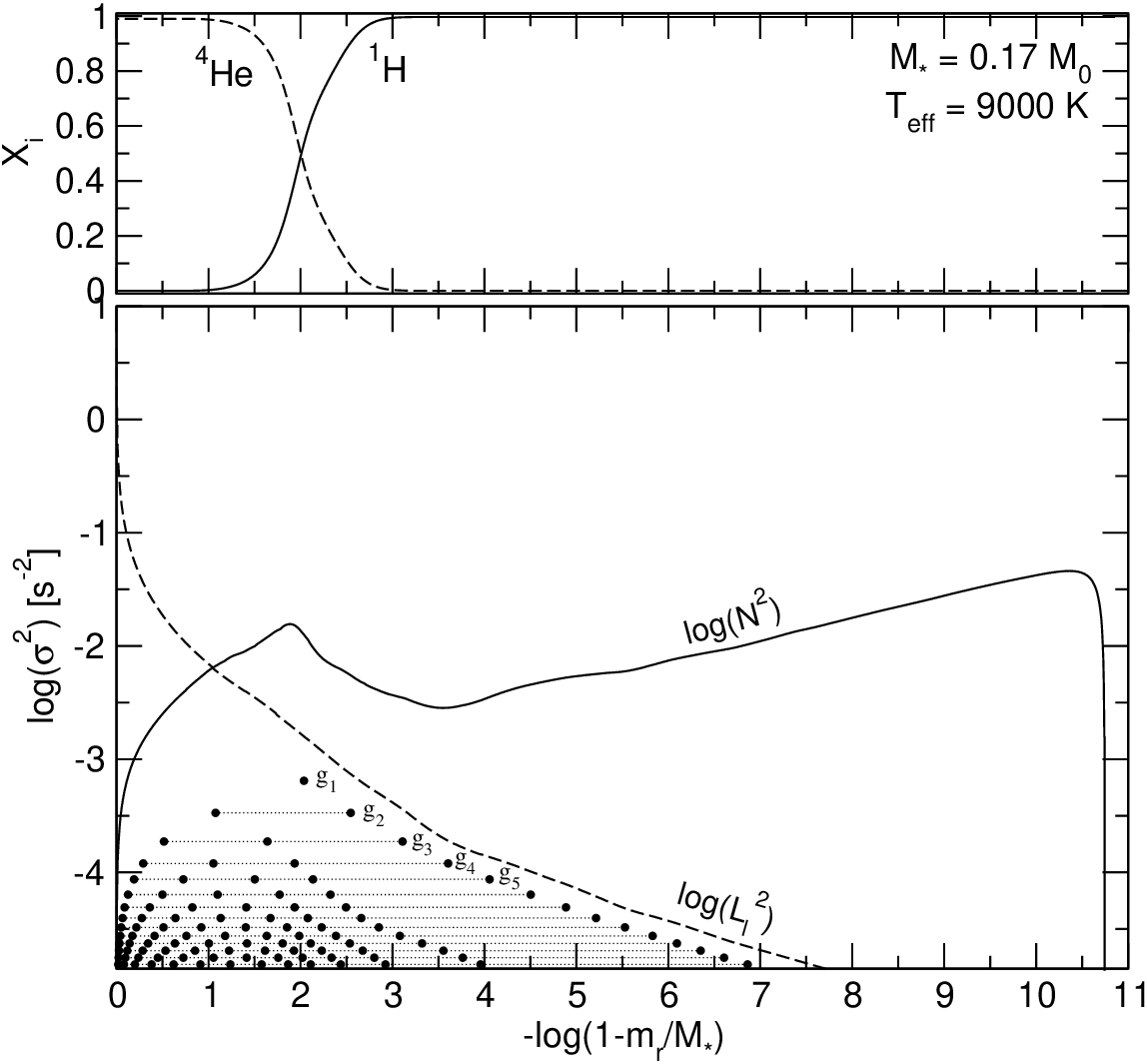}
\caption{The internal chemical profiles of  He and H (upper panel) and
  the propagation diagram  ---the run of the logarithm  of the squared
  critical frequencies---  (lower panel) corresponding to an ELM white
  dwarf  model  of $M_*=  0.17  M_{\odot}$  and  $T_{\rm eff}  \approx
  9\,000$ K.  Dots connected with  thin dotted lines correspond to the
  spatial  location  of  the  nodes  of the  radial  eigenfunction  of
  low-order dipole ($\ell= 1$) $g$-modes.}
\label{propa-0.17}
\end{figure}

\subsection{Asymptotic period spacing}
\label{asymptotic}

For  $g$-modes  with  high   radial  order  $k$  (long  periods),  the
separation  of consecutive  periods ($|\Delta  k|= 1$)  becomes nearly
constant  at a  value  given  by the  asymptotic  theory of  nonradial
stellar  pulsations.   Specifically,  the  asymptotic  period  spacing
(Tassoul et al.  1990) is given by:

\begin{equation} 
\Delta \Pi_{\ell}^{\rm a}= \Pi_0 / \sqrt{\ell(\ell+1)},  
\label{aps}
\end{equation}

\noindent where

\begin{equation}
\label{asympeq}
\Pi_0= 2 \pi^2 \left[\int_{r_1}^{r_2} \frac{N}{r} dr\right]^{-1}.
\end{equation}

\noindent The squared Brunt-V\"ais\"al\"a frequency ($N$) is computed as:  

\begin{equation}
\label{bvf}
N^2= \frac{g^2 \rho}{P}\frac{\chi_{\rm T}}{\chi_{\rho}}
\left[\nabla_{\rm ad}- \nabla + B\right],
\end{equation}

\noindent where the compressibilities are defined as

\begin{equation}
\chi_{\rho}= \left(\frac{d\ln P}{d\ln \rho}\right)_{{\rm T}, \{\rm X_i\}}\ \ \
\chi_{\rm T}= \left(\frac{d\ln P}{d\ln T}\right)_{\rho, \{\rm X_i\}}.
\end{equation}

\noindent The Ledoux term $B$ is computed as (Tassoul et al. 1990):

\begin{equation}
\label{B}
B= -\frac{1}{\chi_{\rm T}} \sum_1^{M-1} \chi_{\rm X_i} \frac{d\ln X_i}{d\ln P}, 
\end{equation}

\noindent where:

\begin{equation}
\chi_{\rm X_i}= \left(\frac{d\ln P}{d\ln X_i}\right)_{\rho, {\rm T}, 
\{\rm X_{j \neq i}\}}.
\end{equation}

\noindent The  expression in Eq.  (\ref{aps}) is  rigorously valid for
chemically  homogeneous  stars.   In   this  equation  (see  also  Eq.
\ref{asympeq}), the dependence on the Brunt-V\"ais\"al\"a frequency is
such that the asymptotic period spacing is larger when the mass and/or
temperature of the model is lower.  This trend is clearly evidenced by
Fig. \ref{asint} for $M_* \gtrsim  0.25 M_{\odot}$, in which we depict
the evolution of  the asymptotic period spacing for  all the sequences
considered in this work.  The higher values of $\Delta \Pi_{\ell}^{\rm
  a}$ for lower  $M_*$ comes from the dependence  $N \propto g$, where
$g$ is the  local gravity ($g\propto M_*/R_*^2$).  On  the other hand,
the  higher values of  $\Delta \Pi_{\ell}^{\rm  a}$ for  lower $T_{\rm
  eff}$ result  from the dependence  $N \propto \sqrt  \chi_{T}$, with
$\chi_{T}  \rightarrow 0$  for increasing  degeneracy  ($T \rightarrow
0$).

\begin{figure}
\centering
\includegraphics[clip, width=0.45\textwidth, angle=0]{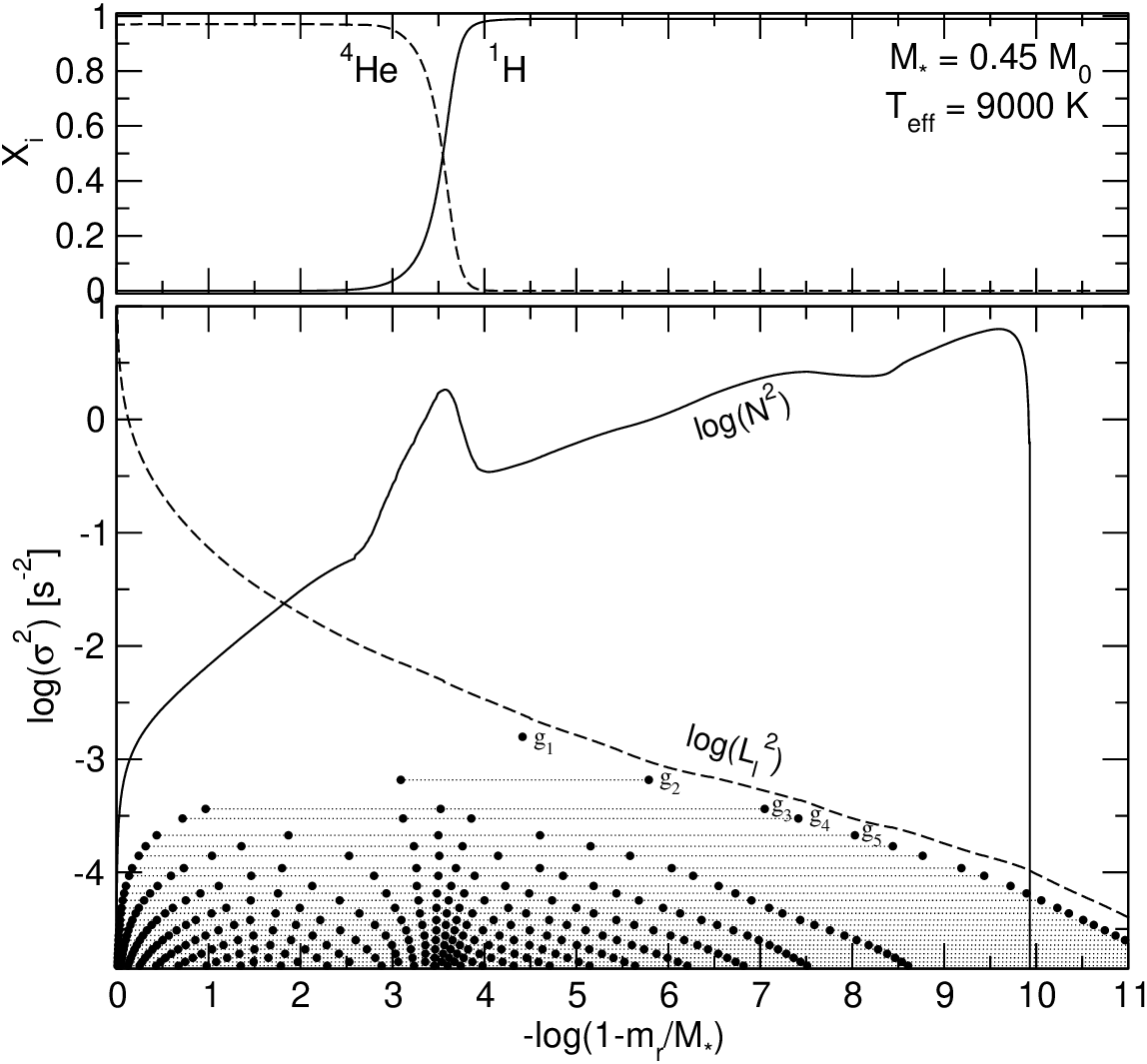}
\caption{Same  as in  Fig.   \ref{propa-0.17}, but  for  a $M_*=  0.45
  M_{\odot}$ white dwarf model.}
\label{propa-0.45}
\end{figure}

Note that,  however, the increase of $\Delta  \Pi_{\ell}^{\rm a}$ with
$T_{\rm eff}$ is  not monotonous for the sequences  with $M_* \lesssim
0.25 M_{\odot}$. In fact, for this range of masses, the period spacing
experiences  a local minimum  before resuming  the final  growth trend
with decreasing effective temperature  according to the predictions of
Eq. (\ref{asympeq}).   That minimum in $\Delta  \Pi_{\ell}^{\rm a}$ is
found at lower  $T_{\rm eff}$ as we go to  lower masses. This somewhat
weird behavior is  induced by the evolving shape  of the He/H chemical
interface.  We postpone  an explanation  for this  feature  of $\Delta
\Pi_{\ell}^{\rm a}$ to Sect. \ref{element_diffusion}.

The strong dependence  of the period spacing on  $M_*$ as evidenced by
Fig. \ref{asint}  could be  used, in principle,  to infer  the stellar
mass  of  pulsating  low-mass   white  dwarfs,  provided  that  enough
consecutive pulsation periods  were available from observations.  Such
a prospect could  be severely complicated by the  fact that the period
spacing of pulsating white dwarfs also depends on the thickness of the
outer envelope  (Tassoul et al.  1990),  where $\Delta \Pi_{\ell}^{\rm
  a}$ is larger  for thinner envelopes.  This is  particularly true in
the context of  low-mass He-core white dwarfs, in  which the models of
ELM objects  harbor H  envelopes that are  several times  thicker than
more   massive  models.    Then,  regarding   the  value   of  $\Delta
\Pi_{\ell}^{\rm a}$, and for a fixed $T_{\rm eff}$, a model with a low
mass and a  thick H envelope could readily mimic  a more massive model
with  a thinner envelope.   We envisage  that, if  a rich  spectrum of
observed  periods were available,  this ambiguity  could be  broken by
including  additional  information of  the  mode trapping  properties,
which   yield   clues   about    the   H   envelope   thickness   (see
Sect. \ref{mode_trapping}).

\subsection{Chemical profiles, critical frequencies and eigenfunctions}
\label{critical_sequences}

Our low-mass white dwarf models are made  of a He core surrounded by a
H outer  envelope.  In  between, there is  a smooth  transition region
shaped  by  the  action   of  microscopic  diffusion,  which  is  self
consistently accounted  for in  LPCODE. In the  upper panels  of Figs.
\ref{propa-0.17} and \ref{propa-0.45} we display the internal chemical
profiles for  He and H corresponding  to two template  models of $M_*=
0.17 M_{\odot}$  and $M_*=  0.45 M_{\odot}$, respectively,  at $T_{\rm
  eff} \approx 9000$  K. The less massive model  is representative of ELM
white dwarfs, and the more  massive model is representative of massive
He-core  white dwarfs ($0.20  \lesssim M_*/M_{\odot}  \lesssim 0.45$).
It  is  worth  emphasizing  two important  differences  between  these
template  models.  First,  the  ELM model  has  a H  envelope that  is
$\approx 35$ times thicker than the massive model. As mentioned in the
Introduction, this  is the result  of the very  different evolutionary
history  of  the  progenitor   stars.   Second,  the  other  notorious
difference   between   the   chemical   structures  shown   in   Figs.
\ref{propa-0.17}  and  \ref{propa-0.45}  is  the  width  of  the  He/H
transition region.  Specifically, this interface is markedly wider for
the ELM  model than  for the massive  one; see Section 3.5 for an 
explanation.

\begin{figure}
\centering
\includegraphics[clip, width=0.45\textwidth, angle=0]{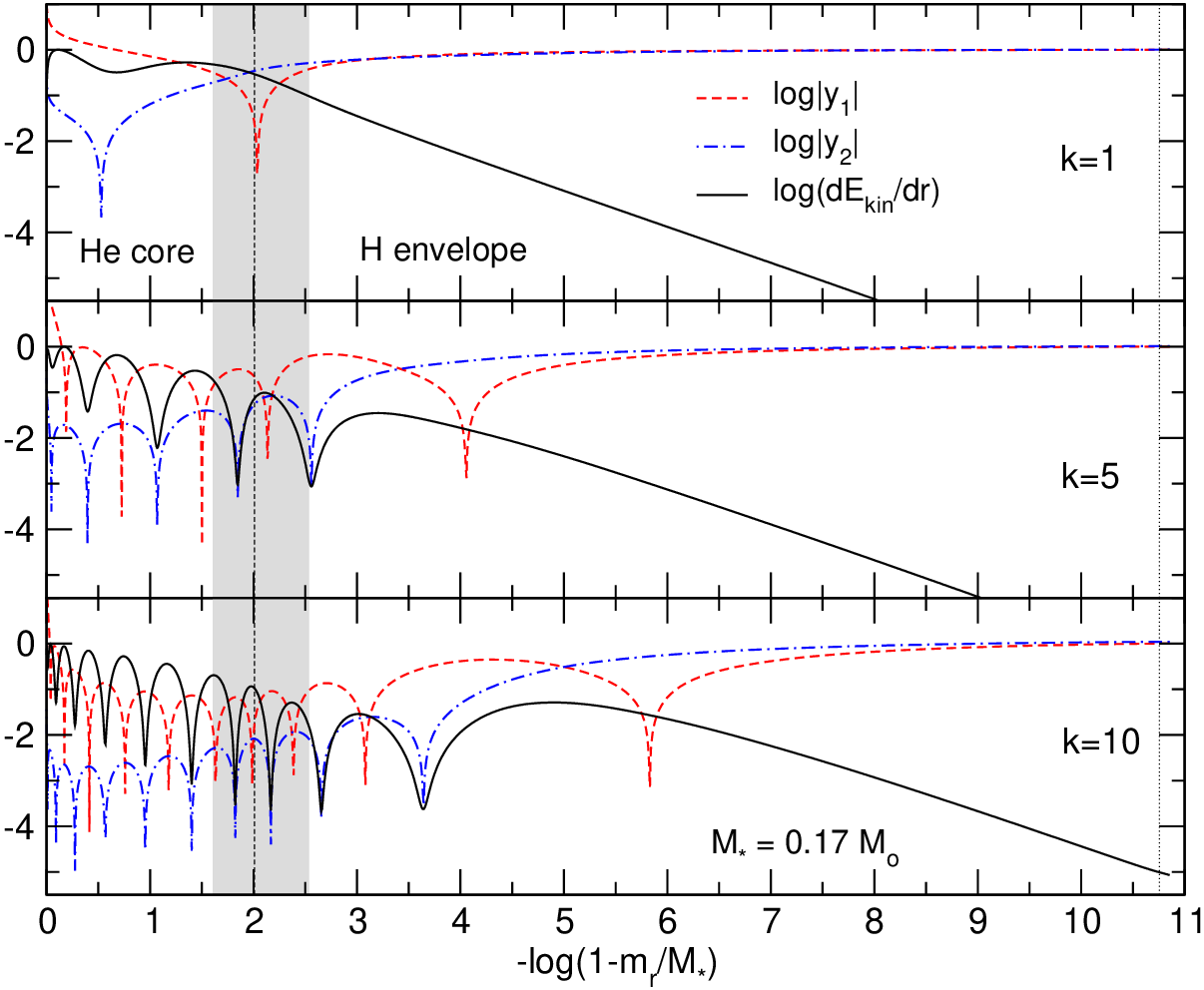}
\caption{The  run  of the  logarithm  of  the  absolute value  of  the
  eigenfunctions $y_1$  and $y_2$, and  the density of  kinetic energy
  $dE_{\rm ekin}/dr$ for $\ell= 1$ modes  with $k= 1$ (upper panel), $k=
  5$ (middle panel),  and $k= 10$ (lower panel),  corresponding to the
  template model with $M_*=  0.17 M_{\odot}$. The vertical dashed line
  and  the associated  shaded  strip  mark the  location  of the  He/H
  chemical  transition  region,   whereas  the  vertical  dotted  line
  indicates the bottom of the outer convective zone.}
\label{autofunciones-0.17}
\end{figure}

The features in  the chemical profiles leave strong  signatures in the
run of the squared critical frequencies, in particular in $N^2$, clearly shown
in the  lower panels of Figs.   \ref{propa-0.17} and \ref{propa-0.45},
which depict  the {\it  propagation diagrams} (Cox  1980; Unno  et al.
1989) corresponding to these models.  The squared Lamb frequency is defined as
$L_{\ell}^2 = \ell (\ell +1)c_{\rm s}^2/r^2$, where $c_{\rm s}$ is the
local adiabatic sound speed.  $g$-modes propagate in the regions where
$\sigma^2  <  N^2,  L_{\ell}^2$,  where $\sigma$  is  the  oscillation
frequency.  Note the  very different shape of $N^2$  for both models. To
begin  with, the squared Brunt-V\"ais\"al\"a  frequency for  the ELM  model is
globally lower than for the  case of the $0.45 M_{\odot}$ model.  This
has to do with the  lower gravity that characterizes the former model,
which  translates into  smaller values  of $N^2$.   Ultimately,  a lower
Brunt-V\"ais\"al\"a  frequency profile  produces a  pulsation spectrum
with longer pulsation periods.

Another  different feature  between the  propagation diagrams  of both
models is related  to the bump at the He/H  transition region. This is
notoriously more narrow and pronounced  for the massive model than for
the ELM  one. As can be  seen in Fig. \ref{propa-0.45},  for the $0.45
M_{\odot}$ model this feature  visibly affects the distribution of the
nodes  of the  radial  eigenfunctions, which  cluster  at the  precise
location of the  peak in $N^2$.  In contrast,  the distribution of nodes
(and so, the shape of the eigenfunctions) is largely unaffected by the
chemical    interface    in    the    case   of    the    ELM    model
(Fig. \ref{propa-0.17}). As we shall see in Sect. \ref{mode_trapping},
this  results in  a weaker  mode trapping  as compared  with  the more
massive models.

\begin{figure}
\centering
\includegraphics[clip, width=0.45\textwidth, angle=0]{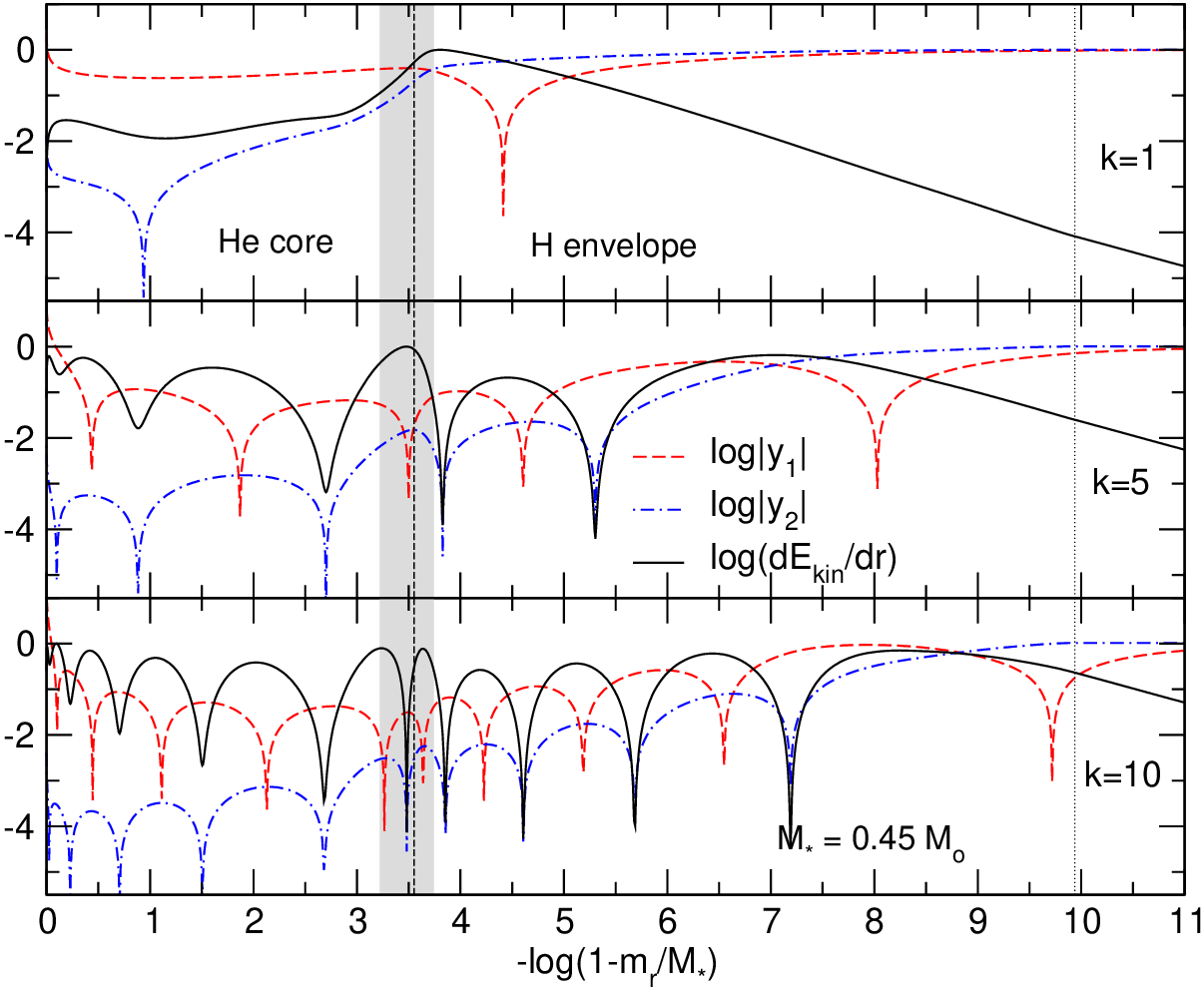}
\caption{Same as Fig. \ref{autofunciones-0.17}, but for the model with
  $M_*= 0.45 M_{\odot}$.}
\label{autofunciones-0.45}
\end{figure}

Finally, another important characteristic of the ELM model is that its
squared  Brunt-V\"ais\"al\"a  frequency exhibits  a  local maximum  at
$-\log(q) \approx  2$, which coincides with  the loci of the  bump due to
the  He/H transition  region.  But  from there,  $N^2$  increases very
slightly outwards, without reaching  much larger values ​​in the surface
layers.  This  is in  contrast with the  case of the  $0.45 M_{\odot}$
model,  in which  the squared  Brunt-V\"ais\"al\"a  frequency exhibits
larger values at the outer layers, resembling the situation in the C/O
core DAV white dwarf models (see Fig.  3 of Romero et al. 2012).  This
quite different shape  of the run of $N^2$  has strong consequences on
the propagation  properties of  eigenmodes.  Specifically, and  due to
the particular shape of $N^2$ (that is, larger in the core than in the
envelope;  see also  Fig.  2 of  SEA10),  the resonant  cavity of  the
eigenmodes  for the  ELM model  is circumscribed  to the  core regions
($-\log  q  \lesssim  2$),  whereas  for  the  $0.45  M_{\odot}$,  the
propagation region extends along the whole model.

This    can    be    appreciated    in   more    detail    in    Figs.
\ref{autofunciones-0.17}  and  \ref{autofunciones-0.45},  in which  we
plot the eigenfunctions $y_1=  \xi_r/r$ and $y_2= (\sigma^2/g)\ \xi_h$
(where  $\xi_r$  and  $\xi_h$   are  the  radial  and  the  horizontal
displacements,  respectively) and the  density of  oscillation kinetic
energy $dE_{\rm kin}/dr$ (see Appendix  A of C\'orsico \& Althaus 2006
for its definition) corresponding to modes with radial order $k= 1, 5$
and $10$ for the two template models ($T_{\rm eff} \approx 9\,000$ K).
From  Fig. \ref{autofunciones-0.17} it  is apparent  that most  of the
spatial oscillations  of the  modes in the  ELM white dwarf  model are
restricted  to the  regions below  the He/H  interface, i.e.,  most of
nodes are located in the region with $-\log(q) \lesssim 2$.  This plot
should be  compared with Fig.  2  of SEA10.  In  summary, $g$-modes in
ELM white dwarfs probe  mainly the stellar core\footnote{However,
  higher-order $g$-modes in the ELM model are less concentrated in the
  core than  low-order ones,  and probe a  significant portion  of the
  star.} and so, they have  an enormous asteroseismic potential, as it
was first recognized by SEA10.

In    contrast,    $g$-modes   in    more    massive   models    (Fig.
\ref{autofunciones-0.45}) oscillate  all along the  stellar structure,
and not just at the  core, although they are more selectively affected
by mode-trapping effects produced by the compositional gradient at the
He/H  interface.  In this  sense, low-mass  He-core white  dwarfs with
$M_*  \gtrsim 0.20  M_{\odot}$ behave  qualitatively similar  to their
massive cousins, the C/O-core DAV white dwarf stars.

\subsection{Mode trapping}
\label{mode_trapping}

\begin{figure}
\centering
\includegraphics[clip, width=0.45\textwidth, angle=0]{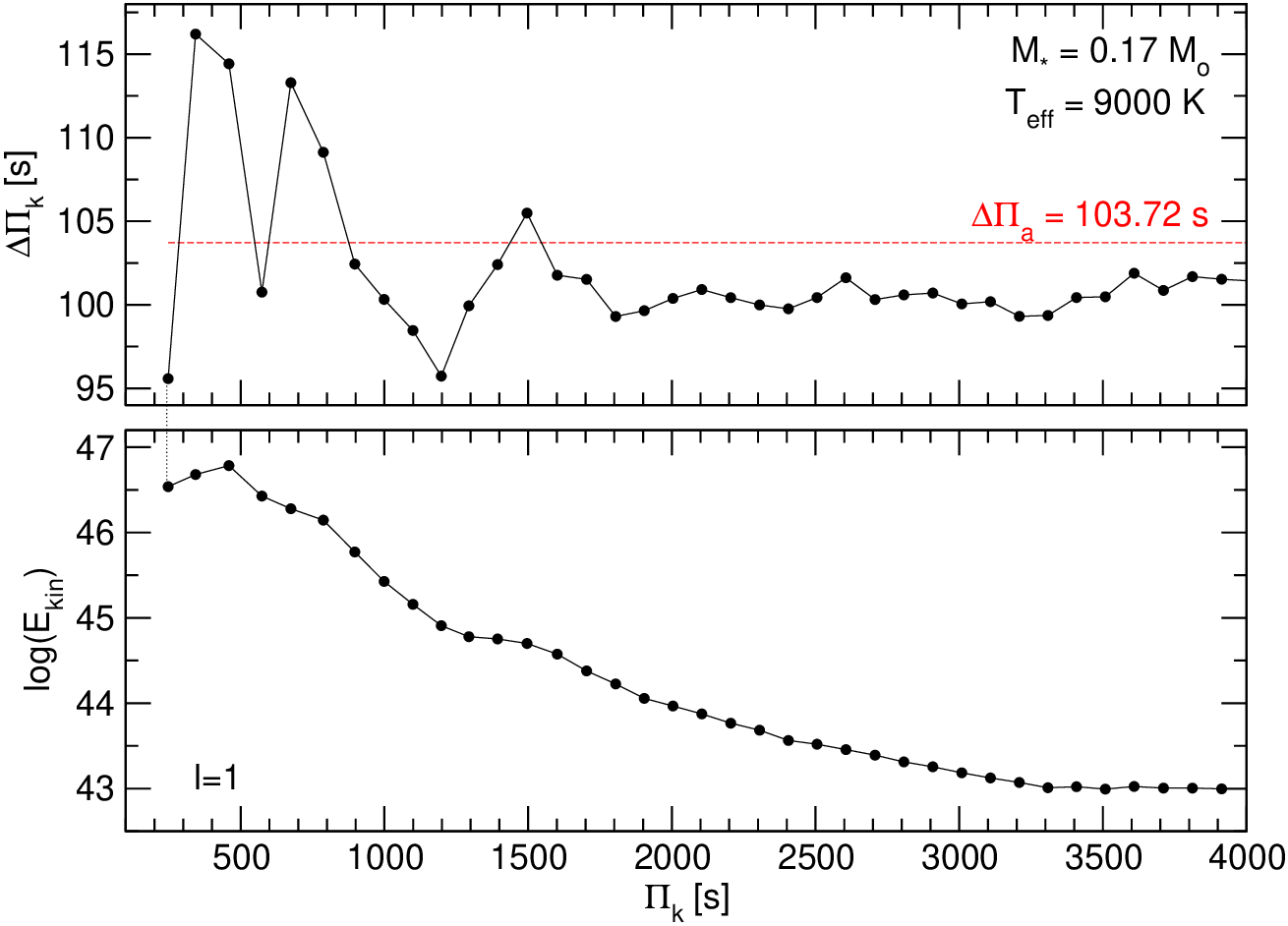}
\caption{The forward period spacing  (upper panel) and the oscillation
  kinetic energy (lower panel) in terms of the ($\ell= 1$) periods for
  the model  with $M_*= 0.17  M_{\odot}$ and $T_{\rm eff}=  9\,000$ K.
  The  red horizontal  line  in  the upper  panel  corresponds to  the
  asymptotic period spacing computed with Eq. (\ref{aps}).}
\label{deltap-ecin-017}
\end{figure}

The  period  spectrum  of  chemically homogeneous  stellar  models  is
characterized by  a constant period separation,  given very accurately
by Eq.   (\ref{aps}).  However, current  evolutionary calculations and
different pieces  of observational evidence indicate  that white dwarf
stars  have  composition  gradients  in their  interiors  (Althaus  et
al. 2010).   The presence of one  or more narrow regions  in which the
abundances of  nuclear species (and  so, the average  molecular weight
$\mu$) are  spatially varying modifies  the character of  the resonant
cavity   in  which   modes   should  propagate   as  standing   waves.
Specifically, chemical interfaces  act like reflecting boundaries that
partially trap  certain modes, forcing  them to oscillate  with larger
amplitudes in specific regions  and with smaller amplitudes outside of
those regions.  The  requirement for a mode to be  trapped is that the
wavelength of its radial  eigenfunction matches the spatial separation
between two interfaces or between one interface and the stellar center
or surface.   This mechanical resonance,  known as mode  trapping, has
been the subject of intense study  in the context of stratified DA and
DB  white dwarf pulsations  (Brassard  et al.   1992,  Bradley et  al.
1993, C\'orsico et  al.  2002a).  In the field of  PG 1159 stars, mode
trapping has  been extensively explored  by Kawaler \&  Bradley (1994)
and C\'orsico \& Althaus (2006).

There are two clear signatures  of mode trapping.  One of them affects
the  distribution of  the oscillation  kinetic energy  ($E_{\rm kin}$)
values in  terms of the  radial order (or  the periods) of  the modes.
Since $E_{\rm  kin}$ is  proportional to the  integral of  the squared
amplitude of eigenfunctions weighted by the density (see Appendix A of
C\'orsico  \& Althaus  2006),  modes propagating  in the  high-density
environment  typical of  the deep  interior of  the star  will exhibit
higher  values than  modes that  are oscillating  in  the low-density,
external regions.  Thus, a  local  maximum in  $E_{\rm kin}$  is usually
associated to  a mode  partially confined to  the core regions,  and a
local  minima corresponds generally to  a mode partially trapped in the
envelope.

\begin{figure}
\centering
\includegraphics[clip, width=0.45\textwidth, angle=0]{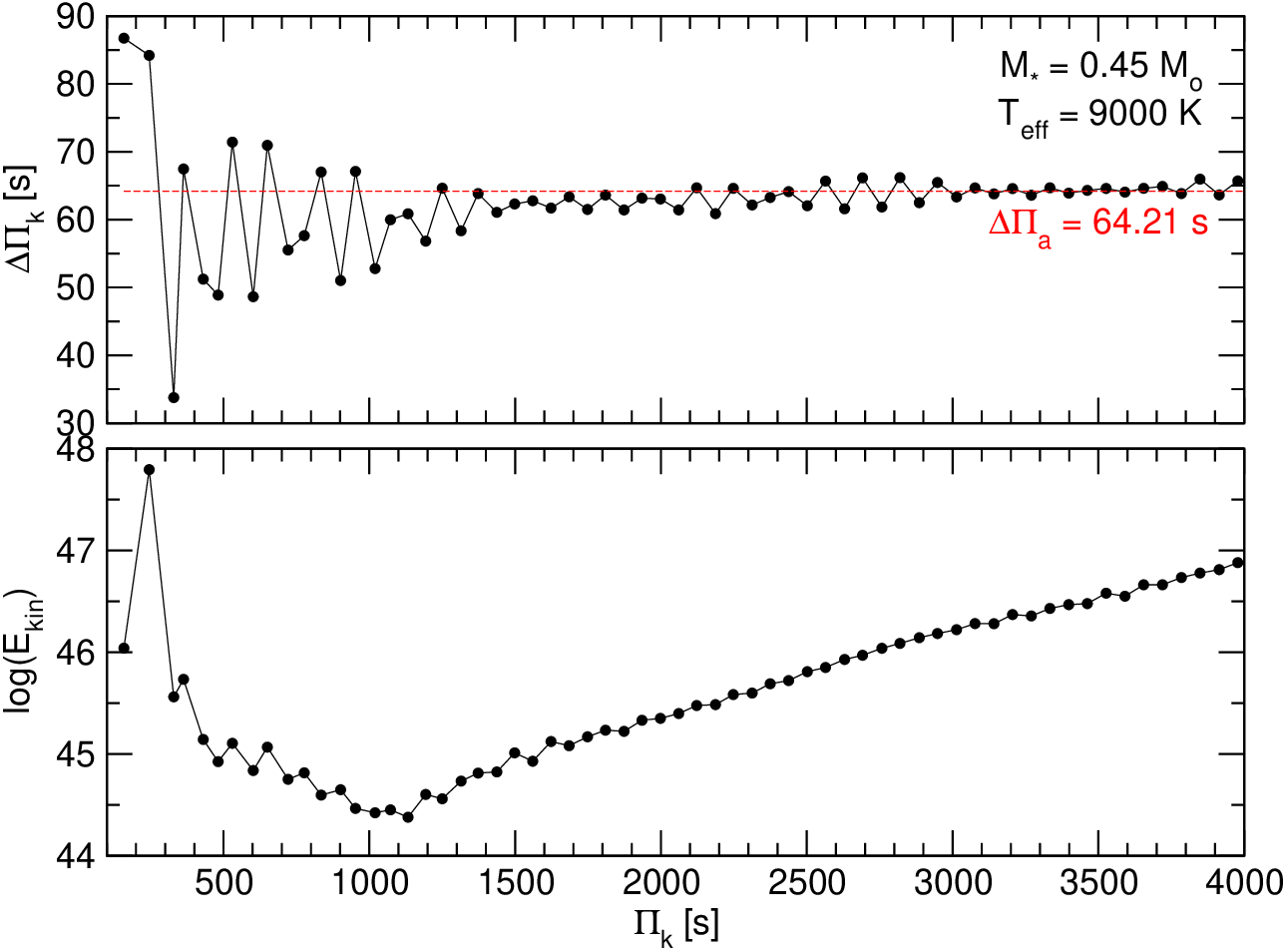}
\caption{Same as Fig. \ref{deltap-ecin-017}, but for the model with 
$M_*= 0.45 M_{\odot}$.}
\label{deltap-ecin-045}
\end{figure}

The second,  and more important  from an observational point  of view,
signature of mode trapping is  that the forward period spacing $\Delta
\Pi_k$  ($\equiv  \Pi_{k+1}-\Pi_k$),  when  plotted in  terms  of  the
pulsation period $\Pi_k$,  exhibits strong departures from uniformity.
It  is the  period difference  between an  observed mode  and adjacent
modes ($\Delta k = \pm 1$) that matters as an observational diagnostic
of mode trapping, at variance  with $E_{\rm kin}$, whose value is very
difficult  to  estimate only  from   observations.  For stellar  models
characterized by  a single  chemical interface, like  the ones  we are
considering here, local minima in $\Delta \Pi_k$ usually correspond to
modes trapped  in the  outer layers, whereas  local maxima  in $\Delta
\Pi_k$ are associated to modes trapped in the core region.

In   the   upper    panels   of   Figs.    \ref{deltap-ecin-017}   and
\ref{deltap-ecin-045}  we  show the  forward  period  spacing for  the
template models with $M_*=  0.17 M_{\odot}$ and $M_*= 0.45 M_{\odot}$,
respectively.  Lower panels depict  the distribution of kinetic energy
for the  same models.   The period spacings  for the  $0.17 M_{\odot}$
model  reach the  asymptotic value  of $103.72$  s for  periods longer
($\approx 5500$ s) than the longest  period shown in the plot. This is
at  variance with what  happens in  the case  of the  $0.45 M_{\odot}$
model, for which $\Delta \Pi_k$  approaches to the asymptotic value of
$64.21$  s  at periods  $\approx  1800$  s.  
The signatures  of  mode
trapping are much  more notorious in the case  of the $0.45 M_{\odot}$
model than for the $0.17 M_{\odot}$ one.  Indeed, mode trapping in the
ELM template  model is appreciable  only in the $\Delta  \Pi_k$ values
for $\Pi_k  \lesssim 1500$  s, and the  $E_{\rm kin}$  distribution is
quite smooth for the complete range of periods shown.  The weakness of
mode-trapping in this model was anticipated in the previous section by
virtue of  the very wide  and smooth He/H chemical  transition region.
In contrast,  for the $0.45  M_{\odot}$ model the  trapping signatures
are   stronger  in  both   the  $\Delta   \Pi_k$  and   $E_{\rm  kin}$
distributions,  although the  kinetic energy  exhibits a  quite smooth
distribution for periods longer than $\approx 2000$ s.

The fact  that the asymptotic period  spacing of the ELM  model is not
very rapidly  reached (in  terms of the  pulsation periods) is  not in
contradiction   with   the   weakness   of   mode   trapping   effects
characterizing  this  model.  This  is  because,  even for  chemically
homogeneous models  ---which completely lack of  mode trapping effects
due to the  absence of chemical gradients--- the  asymptotic regime is
reached  for  modes with  radial  orders  higher  than a  given  limit
(minimum) value of  $k$. This is clearly shown in  Fig. 6 of C\'orsico
\&  Benvenuto (2002), which  displays the  situation for  a chemically
homogeneous white dwarf model made of pure He.

In  summary, our  results  indicate  that mode  trapping  by the  He/H
interface is  important in low-mass  He-core white dwarfs  with masses
$M_* \gtrsim 0.20 M_{\odot}$, but it  is not an important issue in ELM
white dwarfs. This feature severely limits the seismological potential
of these objects to constrain the thickness of the H envelope.

\subsection{Dependence of the pulsations 
with $M_*$ and $T_{\rm eff}$} 
\label{dependence}

\begin{figure}
\centering
\includegraphics[clip, width=0.45\textwidth, angle=0]{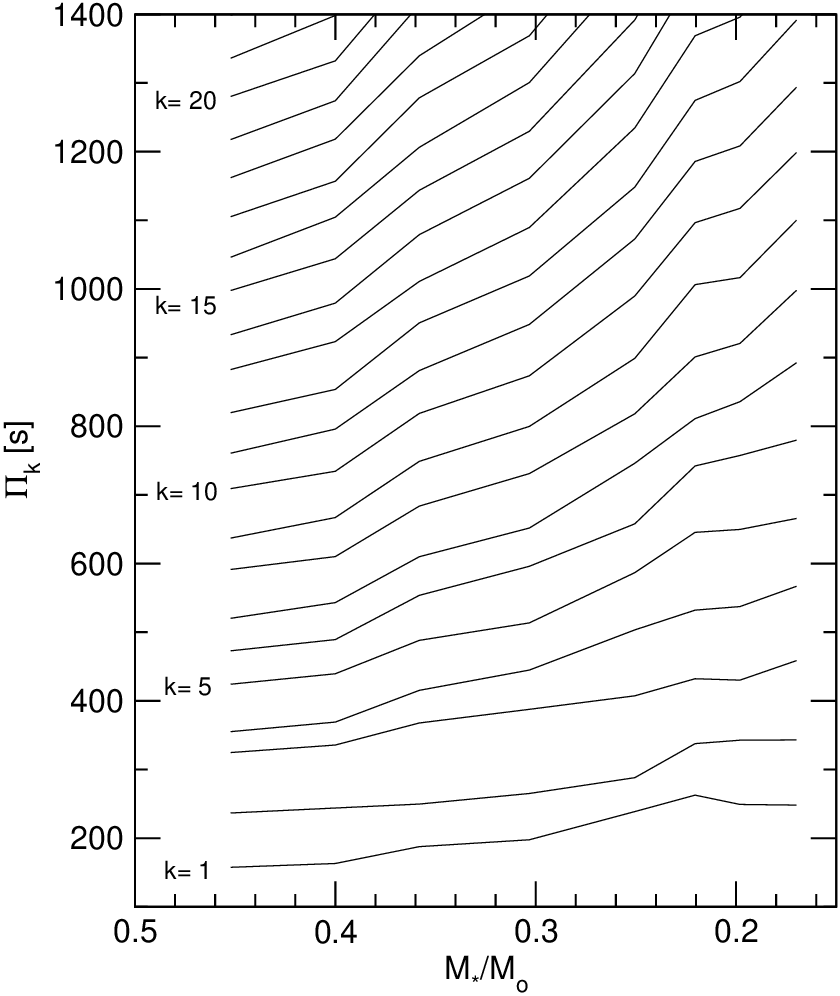}
\caption{Pulsation  periods of  $\ell= 1$  $g$-modes in  terms  of the
  stellar  mass  for $T_{\rm  eff}=  9500$  K.  Periods increase  with
  decreasing $M_*$.}
\label{per-mass}
\end{figure}

\begin{figure}
\centering
\includegraphics[clip, width=0.45\textwidth, angle=0]{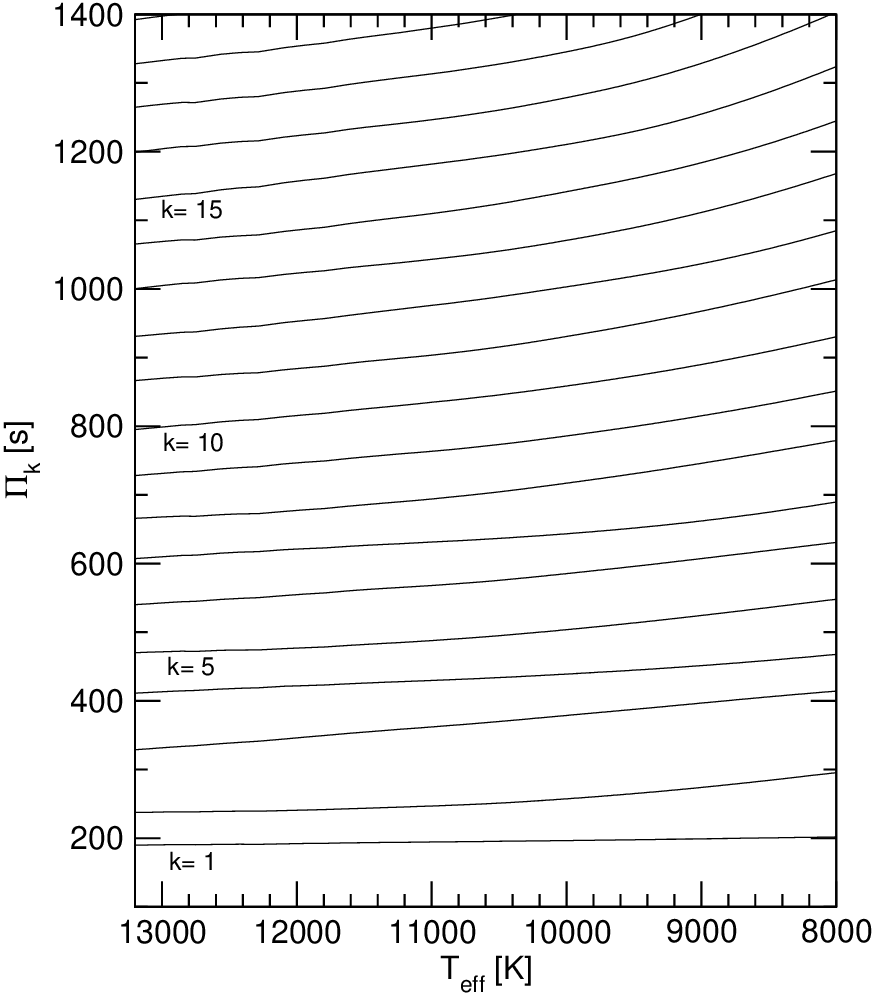}
\caption{Pulsation  periods of  $\ell= 1$  $g$-modes in  terms  of the
  effective  temperature for $M_*=  0.30 M_{\odot}$.  Periods increase
  with decreasing $T_{\rm eff}$.}
\label{per-teff}
\end{figure}

As discussed  in Sect.  \ref{asymptotic},  the period spacing  and the
periods  vary as  the  inverse of  the Brunt-V\"ais\"al\"a  frequency.
This critical  frequency, in turn, increases with  larger stellar mass
and  with higher effective  temperature.  As  a result,  the pulsation
periods  are  longer  for  smaller  mass  (lower  gravity)  and  lower
effective  temperature  (increasing degeneracy).   The  influence of  the
stellar   mass   on  the   pulsation   periods   is   shown  in   Fig.
\ref{per-mass}, where we plot $\Pi_k$  for $\ell= 1$ in terms of $M_*$
for a fixed value of $T_{\rm  eff}$. Note that the increase of $\Pi_k$
with decreasing $M_*$ is less  pronounced for low-order modes than for
higher-order ones. This  is in agreement with the  behavior typical of
the ZZ Ceti  star models (see Bradley 1996).   In Fig.  \ref{per-teff}
we show the evolution of  the pulsation periods with $T_{\rm eff}$ for
the case of models with $M_*= 0.30 M_{\odot}$.  The lengthening of the
periods as  the effective temperature  drops is evident,  although the
effect is less pronounced than  the decrease with the stellar mass (as
compared with Fig. \ref{per-mass}).

\begin{figure}
\centering
\includegraphics[clip, width=0.45\textwidth, angle=0]{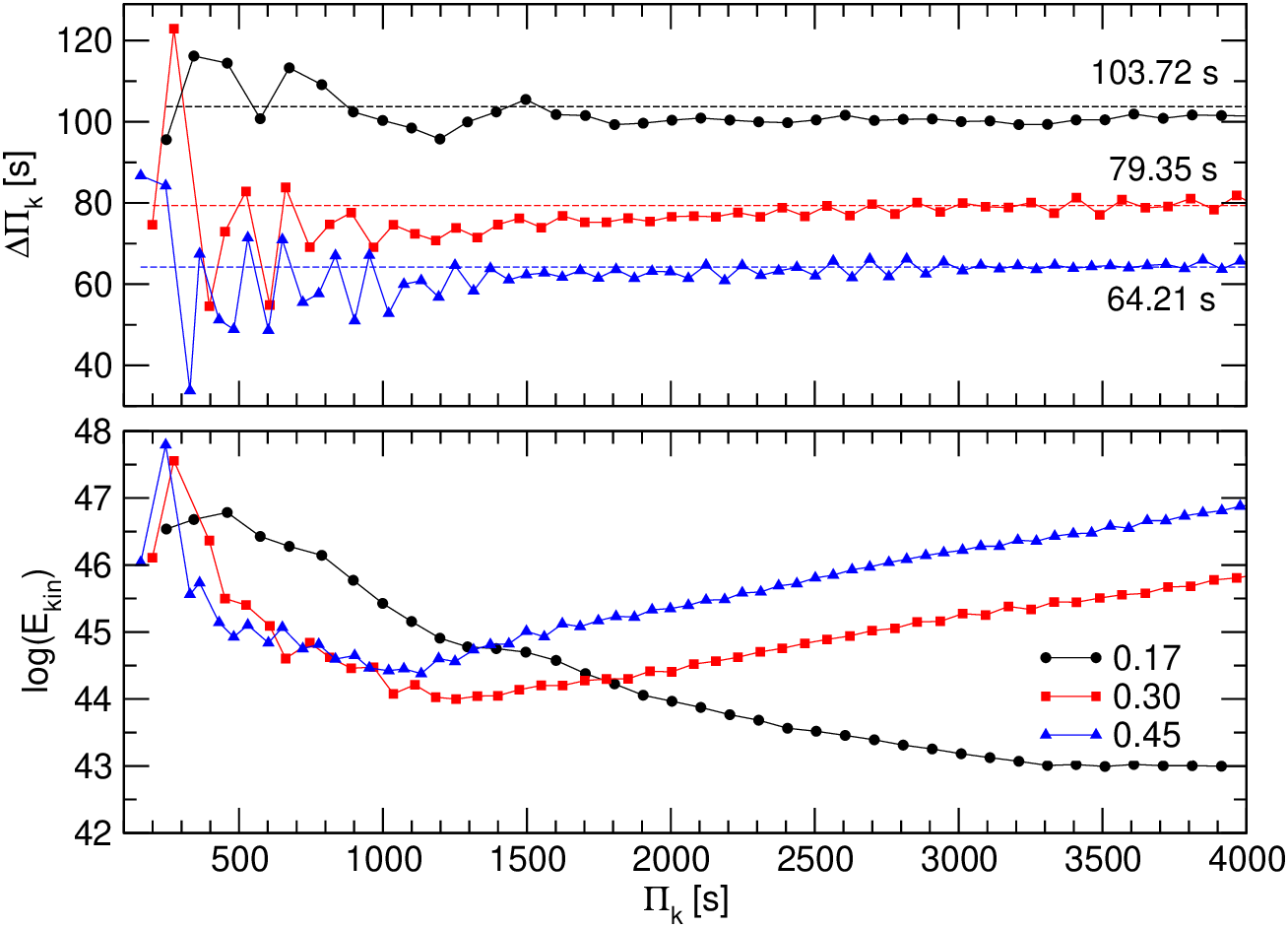}
\caption{The  dipole  forward period  spacing  (upper  panel) and  the
  oscillation kinetic energy (lower panel) in terms of the periods for
  models with $M_*= 0.17  M_{\odot}$, $M_*= 0.30 M_{\odot}$, and $M_*=
  0.45  M_{\odot}$,  all of  them  characterized  by  a $T_{\rm  eff}=
  9\,000$ K. The horizontal lines in the upper panel correspond to the
  asymptotic period spacings computed with Eq. (\ref{aps}).}
\label{ekin-dp-M.eps}
\end{figure}

The effect  of varying $M_*$ on the  period-spacing and kinetic-energy
distributions  for  a  fixed   $T_{\rm  eff}$  is  displayed  in  Fig.
\ref{ekin-dp-M.eps},  in  which  we  present  the  results  for  three
white dwarf models with stellar masses  of $M_*= 0.17, 0.30$ and $0.45
M_{\odot}$ and  the same  effective temperature ($T_{\rm  eff} \approx
9000$ K). As expected, $\Delta  \Pi_k$ strongly changes with $M_*$, as
$\Delta \Pi^{\rm a}$ does. In  fact, the average of the $\Delta \Pi_k$
values varies  from $\approx 60$ s  to $\approx 100$ s  ($\approx 60 \%$)
when  the  stellar  mass   changes  from  $0.45  M_{\odot}$  to  $0.17
M_{\odot}$.    Also,  substantial  changes   in  the   kinetic  energy
distribution are expected when we  vary the stellar mass, as evidenced
by the lower panel of the Fig. \ref{ekin-dp-M.eps}.

\begin{figure}
\centering
\includegraphics[clip, width=0.45\textwidth, angle=0]{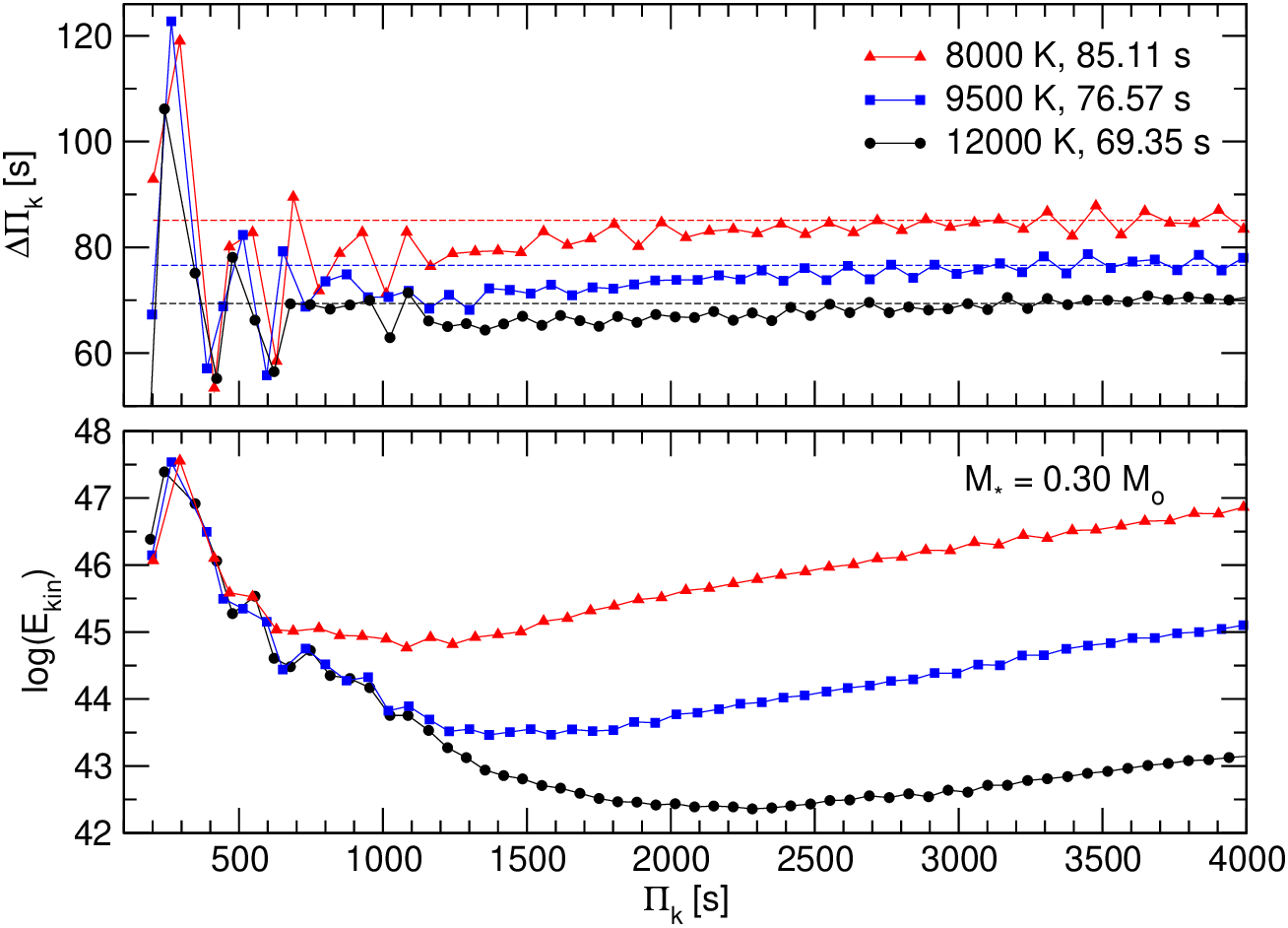}
\caption{Same as Fig. \ref{ekin-dp-M.eps}, but for the case of a model
  with   $M_*=   0.30  M_{\odot}$   and   three  different   effective
  temperatures of $T_{\rm eff}= 12\,000, 9500$ and $8000$ K.}
\label{ekin-dp-Teff}
\end{figure}

Finally, we  examine the effect  of changing $T_{\rm eff}$  on $\Delta
\Pi_k$ and $E_{\rm kin}$  with fixed $M_*$ in Fig. \ref{ekin-dp-Teff},
where we show these quantities in terms of the pulsation periods for a
model with  $M_*= 0.30  M_{\odot}$ and three  different values  of the
effective temperature.  The sensitivity  of the period spacing and the
kinetic  energy on  $T_{\rm  eff}$ are  negligible  for short  periods
($\Pi_k  \lesssim 500$  s)  but  it is  quite  appreciable for  longer
periods. Specifically, the average of the $\Delta \Pi_k$ values varies
from  $\approx  70$ s  to  $\approx  85$ s  ($\approx 20  \%$) when  the
effective temperature changes from $12\,000$ K to $8000$ K.  Thus, the
dependence  of  $\Delta \Pi_k$  with  $T_{\rm  eff}$ is  substantially
weaker  than with  $M_*$.  The  kinetic energy  values also  are quite
sensitive to  changes in  $T_{\rm eff}$, at  least for  periods longer
than  $\approx  500$   s.   For  the  hottest  model   depicted  in  Fig.
\ref{ekin-dp-Teff}  ($T_{\rm eff}=  12\, 000$  K), the  more energetic
modes are those with $k= 1, \cdots, 5$, because they oscillate deep in
the star  where the density is  high.  However, for  the cooler models
($9500$ and  $8000$ K),  the kinetic energy  of the  high-order modes,
which are  characterized by eigenfunctions mainly  concentrated in the
outer  layers, strongly  increase.  This  is because  the  decrease in
$T_{\rm eff}$ produces  a deepening of the outer  convection zone, and
so, the high-order modes feel  gradually the presence of convection as
the  white dwarf  cools down.   Since $g$-modes  in a  convection zone
become evanescent, such modes are  forced to gain larger amplitudes at
regions somewhat  below the base  of the outer convection  zone, where
the  density is  larger. Since  $E_{\rm kin}$  is proportional  to the
integral  of the  squared  eigenfunctions, weighted  by $\rho$,  these
modes oscillate with larger energies.  However, low-order modes remain
rather insensitive to the thickening of the outer convection zone.

\subsection{Effects of element diffusion}
\label{element_diffusion}

\begin{figure}
\centering 
\includegraphics[clip,width=0.45\textwidth, angle=0]{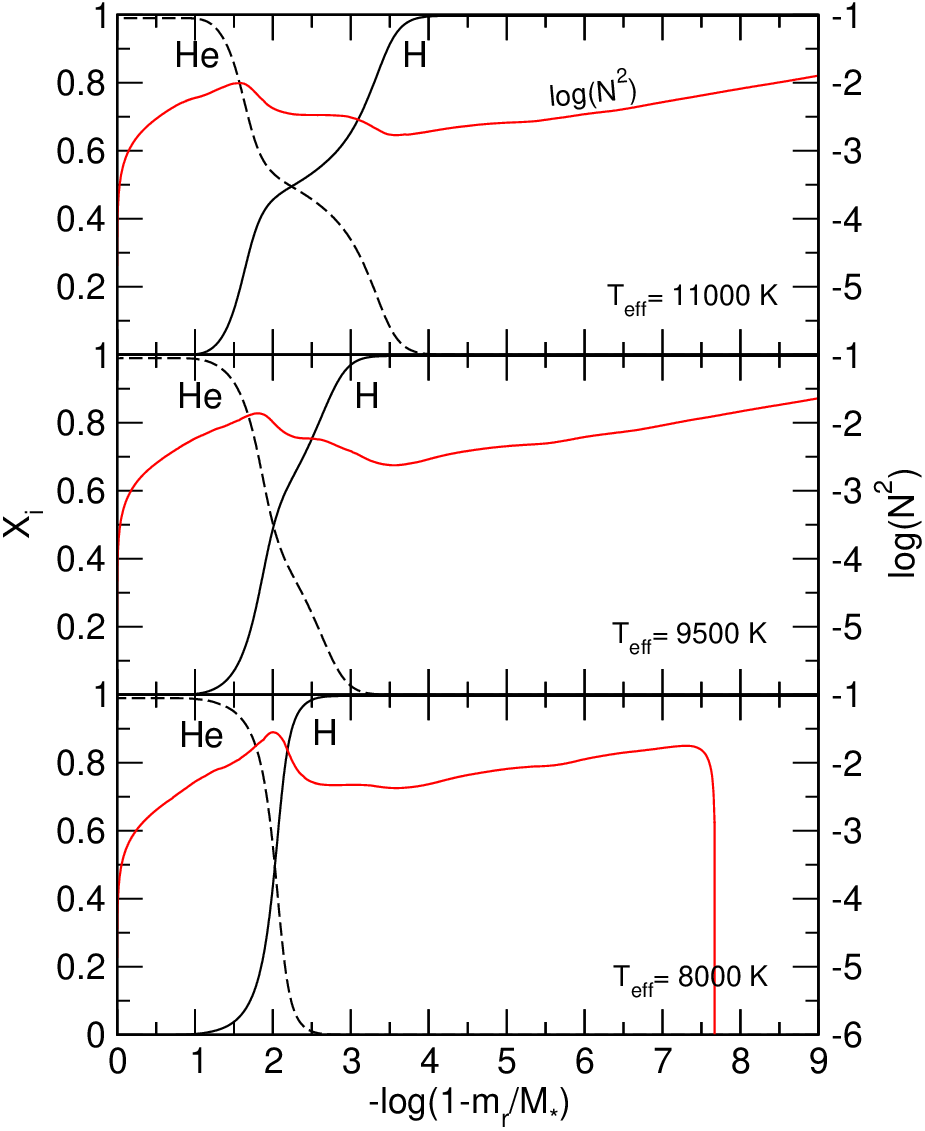}
\caption{The internal chemical profile of  H and He, and the logarithm
  of the squared Brunt-V\"ais\"al\"a frequency, for a model with $M_*=
  0.17  M_{\odot}$  at  different  effective temperatures,  which  are
  indicated in each panel.}
\label{difusion-perfiles}
\end{figure}

Finally, we  discuss the effects that evolving  chemical profiles have
on the  pulsation properties of  low-mass, He-core white  dwarfs.  The
effect  of  element  diffusion  on  DAV stars  has  been  explored  by
C\'orsico et al.  (2002b).  Time-dependent element diffusion, which is
fully taken  into account in  our computations, strongly  modifies the
shape of  the He  and H  chemical profiles as  the white  dwarf cools,
causing H to float to the surface and He to sink down.  In particular,
diffusion  not only  modifies the  chemical composition  of  the outer
layers,  but also  the shape  of the  He/H chemical  transition region
itself.  This is clearly  borne out from Fig.  \ref{difusion-perfiles}
for the  case of  the $0.17 M_{\odot}$  sequence in the  $T_{\rm eff}$
interval ($11\,000-8000$ K).  For  the model at $T_{\rm eff}= 11\,000$
K,   the   H  profile   is   characterized   by  a   diffusion-modeled
double-layered chemical structure, which consists in a pure H envelope
atop an  intermediate remnant  shell rich in  H and He  (upper panel).
This structure still  remains, although to a much  less extent, in the
model at  $T_{\rm eff}= 9500$  K (middle panel).  Finally,  at $T_{\rm
  eff}= 8000$ K, the H profile has a single-layered chemical structure
(lower panel).   Element diffusion processes affect  all the sequences
considered   in   this  paper,   although   the   transition  from   a
double-layered structure  to a single-layered one  occurs at different
effective  temperatures.  The  markedly  lower  surface  gravity  that
characterizes  the less  massive model,  results in  a less  impact of
gravitational settling, and eventually  in a wider chemical transition
in the  model with $M_*= 0.17  M_{\odot}$.  Because of  this fact, the
sequences  with larger  masses reach  the single-layered  structure at
higher effective temperatures.

The changes  in the  shape of the  He/H interface are  translated into
non-negligible  changes  in  the  profile of  the  Brunt-V\"ais\"al\"a
frequency,  as can  be  appreciated in  the  plot.  In  fact, at  high
effective  temperatures, $N^2$  is characterized  by two  bumps, which
merge  into  a  single one  when  the  chemical  profile at  the  He/H
interface  adopts  a  single-layered  structure.  The  fact  that  the
chemical profiles evolve  with time in our models  is at variance with
the  claim by SEA10,  who argue  that in  their ELM  models, diffusive
equilibrium is  valid in the He/H  transition region. In  the light of
our results, we conclude that this assumption is wrong.

In order to  quantify the impact of the  diffusively evolving chemical
profiles on the  period spectrum of low-mass He-core  white dwarfs, we
focus    on    the    same    ELM   sequence    analyzed    in    Fig.
\ref{difusion-perfiles}.    Specifically,   we   have  considered   an
additional  sequence  of  models  for  which we  switch  off  chemical
diffusion in  the LPCODE evolutionary  code from $T_{\rm  eff} \approx
11\,000$  K downwards.   This  allow  us to  isolate  the effect  that
diffusion processes induce on the pulsation periods from those changes
induced by cooling.  We found  substantial changes in the value of the
periods when  diffusion is neglected,  depending on the  specific mode
considered and the value  of the effective temperature.  For instance,
for the  $k= 3$ mode ($\Pi_3  \approx 460$ s) at  $T_{\rm eff} \approx
9000$ K, a variation of $5 \%$ in the value of the period is expected.

We  conclude that  time-dependent  element diffusion  does affect  the
pulsation spectrum of low-mass and  ELM white dwarfs, and that it must
be  taken into  account in  future pulsational  analysis of  ELM white
dwarfs.

We close this  section with an explanation to  the behavior of $\Delta
\Pi_{\ell}^{\rm  a}$ in  terms of  the effective  temperature  for the
sequences with  $M_* \leq 0.22 M_{\odot}$, shown  in Fig.  \ref{asint}
(Sect.  \ref{asymptotic}).    As  we  have   seen,  element  diffusion
processes lead to the  formation of a double-layered configuration for
the  H profile  in our  models.   This configuration  consists in  two
regions  where the  He  and  H abundances  exhibit  a steeped  spatial
variation,   which  translates   into  two   local  features   in  the
Brunt-V\"ais\"al\"a  frequency.   As  evolution  proceeds,  the  outer
interface shifts  downwards, gradually getting closer  to the internal
interface. The merging  of the two bumps in $N$  tends to increase the
integrand of  Eq.  (\ref{asympeq}), thus  producing a decrease  in the
asymptotic period  spacing.  Thus, for a given  $T_{\rm eff}$, $\Delta
\Pi_{\ell}^{\rm    a}$   reaches   a    local   minimum    (see   Fig.
\ref{asint}).  Later,   cooling  again  dominates   and  then  $\Delta
\Pi_{\ell}^{\rm  a}$  increases,  as  expected.   We  found  that  the
effective temperature at which $\Delta \Pi_{\ell}^{\rm a}$ reaches the
minimum is higher for larger stellar masses. Thus, this minimum is not
seen  in Fig.   \ref{asympeq} for  the sequences  with $M_*  \geq 0.25
M_{\odot}$ because in  these cases that minimum lies  out of the range
of $T_{\rm eff}$ considered.

\section{Nonadiabatic computations and the case of SDSS J184037.78$+$642312.3} 
\label{sdss}

Having  thoroughly described  in  the previous  sections the  $g$-mode
adiabatic  seismic  properties of  our  low-mass  He-core white  dwarf
models, in what follows we will focus our attention on the only (up to
now)  ELM  white  dwarf  discovered  to  be  a  pulsating  star,  SDSS
J184037.78+642312.3.   This  DA  white  dwarf, which  is  the  coolest
($T_{\rm eff}=  9100 \pm 170$)  and the lowest-mass  ($M_*\approx 0.17
M_{\odot}$)  known pulsating ($  1200 \lesssim  \Pi \lesssim  4500$ s)
object of  this kind, was discovered  by Hermes et al.   (2012) in the
context of a systematic search for variable He-core white dwarfs among
the large number of ELM white dwarfs discovered through the ELM Survey
(Brown et al.   2010, 2012; Kilic et al.   2011, 2012).  The discovery
of SDSS J184037.78+642312.3 opens the  possibility to use the tools of
white dwarf asteroseismology to  obtain valuable information about the
internal structure of low-mass white dwarf stars.

Here, we do  not attempt an asteroseismological fitting  to this star,
because the period data are  still far from be definitive.  Rather, we
concentrate  mainly on the  question of  whether our  low-mass He-core
white dwarf  models are able  to predict  the pulsations  exhibited by
SDSS J184037.78+642312.3  at the right  effective temperature, stellar
mass and range of observed periods.

\begin{figure}
\centering
\includegraphics[clip, width=0.45\textwidth, angle=0]{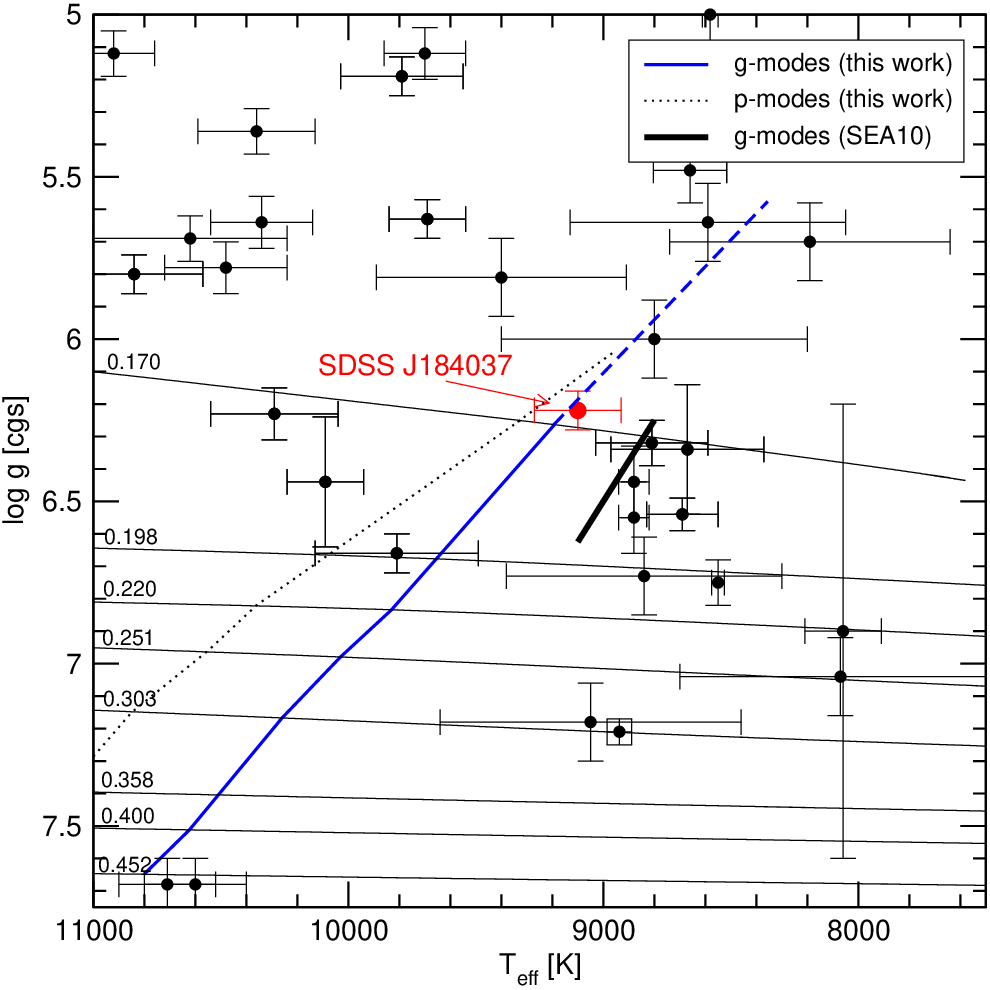}
\caption{A  zoom  of the  $T_{\rm  eff} -  \log  g$  diagram shown  in
  Fig. \ref{tracks} showing the region where is located the only known
  (so far) pulsating ELM white dwarf, SDSS J184037.78+642312.3 (Hermes
  et  al.   2012)  displayed  with  a  red  circle.   We  include  the
  approximate $\ell= 1$ $g$-mode  blue edge  of the theoretical  ELM instability
  domain derived  by SEA10 (thick black solid line), 
  and the  $\ell= 1$ $g$-mode blue
  edge computed in  this work (blue solid line)  for all the sequences
  considered,  included that  of $M_*=  0.17 M_{\odot}$,  and extended
  towards  lower  gravities  (blue dashed line). We  also  show  our
  $\ell= 1$ $p$-mode blue edge of instability (thin dotted line).}
\label{tracks-zoom}
\end{figure}

In order  to investigate the plausibility of  excitation of pulsations
in  our  models, we  performed  a  linear  stability analysis  on  our
complete  set  of  evolutionary  sequences.   We found  that  a  dense
spectrum  of $g$-modes  are excited  by  the $\kappa-\gamma$-mechanism
acting in the H partial ionization zone for all the masses considered,
and that there exists a  well-defined blue (hot) edge of instability of
He-core white dwarfs, which is the low-mass analog to the blue edge of
the  ZZ  Ceti instability  strip.   We  also  found numerous  unstable
$p$-modes;  the corresponding blue  edge is  hotter than  the $g$-mode
blue edge and has a lower  slope. In Fig.  \ref{tracks-zoom} we show a
zoom   of  the   Fig.    \ref{tracks}  at   the   region  where   SDSS
J184037.78+642312.3 is  located. We have included the location of 
the theoretical  dipole ($\ell= 1$) blue edges of the instability 
domain of low-mass white 
dwarfs. The  $g$-mode blue edge  computed in
this work is marked with a blue solid line, and it is extended towards
lower gravities  with a dashed blue  line.  The $p$-mode  blue edge is
displayed   with   a    thin   dotted   line.    Interestingly,   SDSS
J184037.78+642312.3 is slightly cooler  than the derived $g$-mode blue
edge,  and  so,   our  computations  \emph{does  predict}  long-period
pulsations in  this star.   It should be  kept in mind,  however, that
this result  could change if  a MLT prescription with  less convective
efficiency  were  assumed.   Our  stability  analysis  relies  on  the
frozen-convection  approximation that  is known  to  give satisfactory
predictions for the location of the $g$-mode blue edge of other pulsating white
dwarfs (DAVs,  see Brassard \&  Fontaine 1999; DBVs, see  C\'orsico et
al.  2009).  Recently, a detailed  study by van Grootel et al.  (2012)
showed that the blue edge  of DAV models with $0.6 M_{\odot}$ computed
with the frozen convection approximation does not dramatically differ
from that  obtained with a time-dependent convection  treatment in the
nonadiabatic computations.  We note that also  $p$-modes are predicted
by  our  analysis  to  be destabilized  in  SDSS  J184037.78+642312.3,
although  no  short-period  luminosity  variations  characteristic  of
acoustic modes have been detected in this star so far.

\begin{figure}
\centering
\includegraphics[clip, width=0.50\textwidth, angle=0]{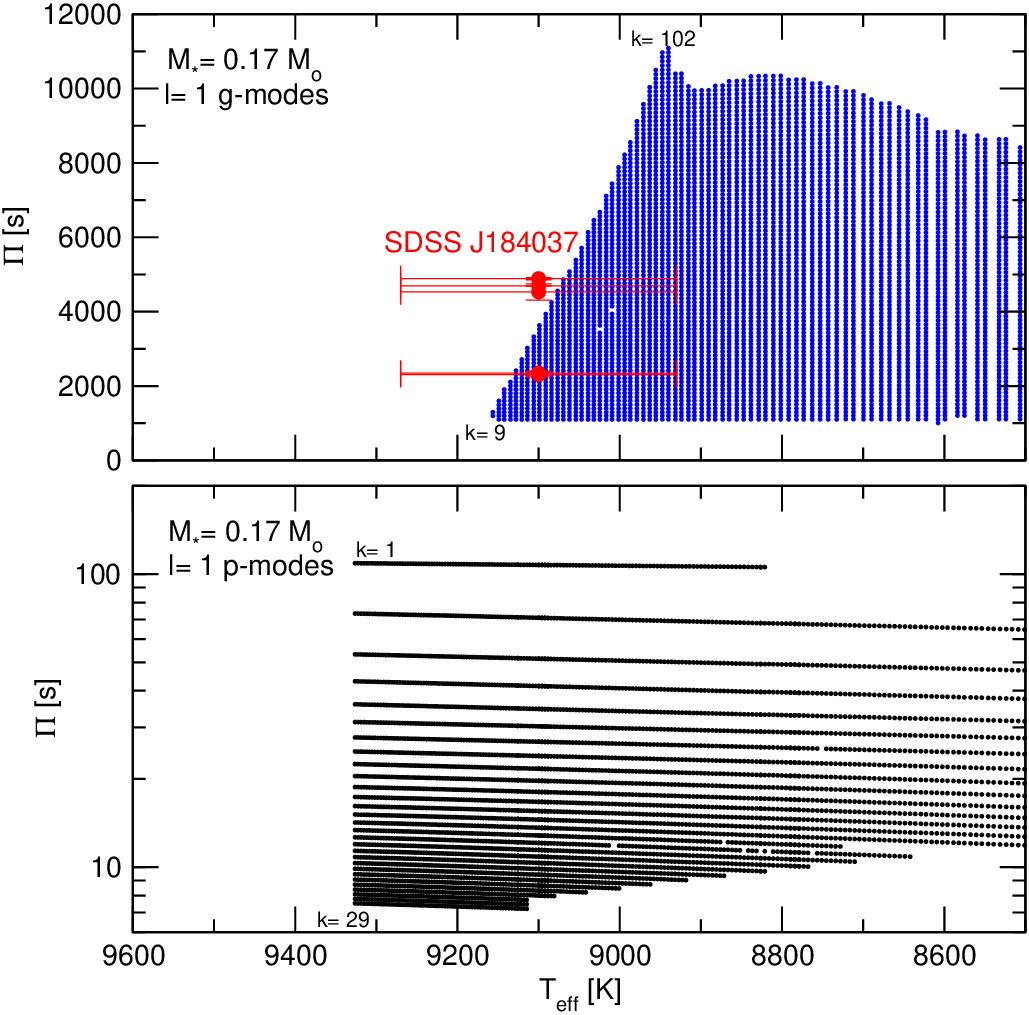}
\caption{Upper panel:  the instability domain  (blue dots) on
  the $T_{\rm eff}-\Pi$ plane for $\ell= 1$ $g$-modes corresponding to
  our set of  models with $M_* = 0.17 M_{\odot}$.   Also shown are the
  periodicities measured in  SDSS J184037.78+642312.3 (red points with
  error bars).  Lower panel: same as upper panel, but for the $p$-mode
  instability domain (black dots).}
\label{noad}
\end{figure}

We  emphasize that,  besides  SDSS J184037.78+642312.3,  our  nonadiabatic
analysis  predicts  $g$- and $p$-mode pulsational  instabilities  in the  other
stars at the right of the blue edges. We believe that it would be worthwhile
to examine  photometrically these stars  in order to see  whether they
are pulsating or not.

\begin{table}
\centering
\caption{$g$-mode pulsation   quantities   of  the   ELM   white  dwarf   model
  representative  of  the  pulsating  star  SDSS  J184037.78+642312.3,
  characterized by $M_*= 0.17 M_{\odot}$, $T_{\rm eff}= 9099$ K, $\log
  g=  6.28$,  and $\log(L/L_{\odot})= 0.0156$.}
\begin{tabular}{ccccc}
\hline
\hline
\noalign{\smallskip}
 $k$  & $\Pi_k$ & $\Delta \Pi_k$ & $d\Pi_k/dt $ & $\log E_{\rm kin}$ \\
      & $[{\rm s}]$ & $[{\rm s}]$ & $[10^{-16} {\rm s/s}]$ & $[{\rm erg}]$  \\
\hline
\noalign{\smallskip}
  1 & 249.50 &      93.29  &   1.88    &  46.49    \\     
  2 & 342.79 &     116.73  &   0.16   &  46.68    \\    
  3 & 459.52 &     112.23  &   3.54    &  46.75    \\    
  4 & 571.75 &     100.36  &   10.23   &  46.39    \\    
  5 & 672.10 &     114.66  &   11.66   &  46.30    \\    
  6 & 786.77 &     109.52  &   11.13   &  46.17    \\    
  7 & 896.29 &     103.81  &   6.09    &  45.80    \\    
  8 & 1000.11 &    100.69  &   4.97    &  45.45    \\    
  9 & 1100.79 &     98.68  &   4.85    &  45.15    \\    
 10 & 1199.47 &     96.00  &   5.10    &  44.88    \\    
 11 & 1295.48 &     98.84  &   6.07    &  44.72    \\    
 12 & 1394.31 &    102.87  &   8.31    &  44.67    \\    
 13 & 1497.19 &    104.76  &    10.17   &  44.62    \\    
 14 & 1601.95 &     103.29  &   9.71    &  44.48    \\    
 15 & 1705.23 &     102.36  &   9.08    &  44.31    \\    
 16 & 1807.60 &     99.646  &   7.41    &  44.10    \\    
 17 & 1907.24 &     100.88  &   9.49    &  43.95    \\    
 18 & 2008.12 &     100.55  &   6.64    &  43.80    \\    
 19 & 2108.68 &     101.61  &   8.15    &  43.70    \\    
 20 & 2210.29 &     101.45  &   7.43    &  43.57    \\    
 21 & 2311.74 &     101.16  &   9.14    &  43.47    \\    
 22 & 2412.90 &     100.94  &   6.54    &  43.34    \\    
 23 & 2513.84 &     100.89  &   7.67    &  43.25    \\    
 24 & 2614.74 &     101.62  &   6.15    &  43.15    \\    
 25 & 2716.36 &     102.42  &   9.29    &  43.08    \\    
 26 & 2818.77 &     101.40  &   9.21    &  42.98    \\    
 27 & 2920.18 &     102.07  &   7.63    &  42.89    \\    
 28 & 3022.25 &     100.59  &   7.40    &  42.79    \\    
 29 & 3122.84 &     102.15  &   7.17    &  42.70    \\    
 30 & 3224.99 &     100.49  &   7.64    &  42.62    \\    
 31 & 3325.48 &     100.94  &   6.18    &  42.53    \\    
 32 & 3426.42 &     100.50  &   5.37    &  42.47    \\    
 33 & 3526.92 &     102.39  &   5.32    &  42.44    \\    
 34 & 3629.32 &     102.39  &   10.02   &  42.42    \\    
 35 & 3731.71 &     103.03  &   7.97    &  42.38    \\    
 36 & 3834.75 &     102.31  &   8.05    &  42.34    \\    
 37 & 3937.05 &     103.63  &   9.80    &  42.30    \\    
 38 & 4040.69 &     102.87  &   11.09   &  42.27    \\    
 39 & 4143.56 &     102.58  &   8.82    &  42.21    \\    
 40 & 4246.14 &     102.66  &   11.93   &  42.18    \\    
 41 & 4348.80 &     102.79  &   9.66    &  42.13    \\    
 42 & 4451.59 &     102.89  &   14.30   &  42.12    \\    
 43 & 4554.47 &     103.09  &   13.62   &  42.10    \\    
 44 & 4657.56 &     103.53  &   16.79  &  42.10    \\  
 45 & 4761.09 &     103.67  &   16.37  &  42.09   \\    
 46 & 4864.77 &     103.99  &   22.30  &  42.11   \\
 47 & 4968.76 &     105.33  &   21.29  &  42.11   \\
\noalign{\smallskip}
\hline
\hline
\end{tabular}
\label{tabla2}
\end{table}

The thick black solid line in Fig. \ref{tracks-zoom} corresponds
to  the  $g$-mode  blue  edge  of  the  theoretical  ELM  white  dwarf
instability region derived by SEA10 by using the instability criterion
$\Pi \leq 8 \pi\ \tau_{\rm th}$ (Brickhill 1991; Wu \& Goldreich 1999)
satisfied by $\ell= 1$ and $k=  1$ modes, where $\tau_{\rm th}$ is the
thermal timescale at the base  of the convection zone.  This criterion
has been shown to be accurate  in the case of ZZ Ceti pulsators, which
are characterized  by significantly  higher gravities, but  SEA10 were
forced to extrapolate its predictions to the low-gravity regime of the
ELM white dwarfs. The blue edge  of SEA10 is about $350$ K cooler than
ours at  $\log g  \approx 6.2$,  and in addition,  its slope  is somewhat
larger,   most  likely  the   result  of   their  using   a  different
parametrization   of  convective   efficiency    ($\alpha=  1.0$,
  Steinfadt  2011).  At  variance  with our  results,  the blue  edge
derived by  SEA10 fails to  predict pulsational instabilities  in SDSS
J184037.78+642312.3.   We note  that, in  order to  estimate  the blue
edge, SEA10 implicitly  \emph{assume} that the $k= 1$,  $\ell= 1$ mode
must  be unstable,  something that  is  not guaranteed  at the  outset
without performing  detailed nonadiabatic pulsation  computations (see
below).  If  the blue  edge were derived  by considering  higher order
modes ($k  > 1$),  then the discrepancy  would be even  larger because
their blue edge should shift to lower effective temperatures.

We  explored the  domain  of unstable  dipole  modes in  terms of  the
effective temperature for the  sequence of $M_*= 0.17 M_{\odot}$.  For
this sequence, $g$-modes become unstable at $T_{\rm eff} \approx 9200$
K.   Fig. \ref{noad}  show the  instability domains  of both  $g$- and
$p$-modes on the $T_{\rm eff}-\Pi$ diagram.  In the case of $g$-modes,
we  found that  unstable modes  have radial  orders $k  \geq  9$ ($\Pi
\gtrsim  1100$ s), where  the most  unstable modes  for each  value of
$T_{\rm eff}$ have  the longest period.  Modes with $k=  6, 7$ and $8$
are  only  marginally   unstable  (their  stability  coefficients  are
extremely  small) and  have not  been included  in the  plot. Finally,
modes with $k= 1, \cdots, 5$ are pulsationally stable.  In the case of
$p$-modes, we  found that these  modes destabilize at  somewhat higher
effective  temperatures ($T_{\rm  eff} \approx  9330$ K).  At variance
with  the case of  $g$-modes, low-order  $p$-modes with  $k= 1,  2, 3,
\cdots, 29$ ($  109 \gtrsim \Pi \gtrsim 7.5$  s) simultaneously become
unstable  at the  blue edge,  being  the most  unstable ones  (largest
stability  coefficients)  those with  the  shortest  periods. For  the
unstable  models,  the  asymptotic  frequency  spacing  of  $p$-modes,
defined as (Unno et al. 1989):

\begin{equation} 
\Delta \nu^{\rm a} = \left( 2 \int_0^R \frac{dr}{c_S}\right)^{-1},
\end{equation}

\noindent ranges from $\Delta \nu^{\rm  a}= 4.67$ mHz (at $T_{\rm eff}
\approx 9330$  K) to $\Delta \nu^{\rm  a}= 6.20$ mHz  (at $T_{\rm eff}
\approx 8000$ K) for our $M_*= 0.17 M_{\odot}$ sequence.

 We  have  completed a  reanalysis  of  the  periods detected  in  SDSS
J184037.78+642312.3  using nearly 33.0  hr of  photometric observation
from October 2011 to July 2012. Our observing and reduction techniques
are identical to those described  in Hermes et al. (2012), but include
more than 27.0 hr of additional observations. Aliasing of the data has
made  determining the periods  especially difficult,  so we  have made
every  attempt  to  be  as  conservative  as  possible  in  our  error
estimates. We include in Fig.  \ref{noad} the periods detected in SDSS
J184037.78+642312.3   as    red   dots   with    their   corresponding
uncertainties.

The highest-amplitude mode  excited in the star occurs  at $4697.8 \pm
4.3$ s ($51  \pm 5$ mma); this periodicity appears in  the 5.5 hr of
October 2011 discovery data, but was the second-highest alias of the
4445 s  mode quoted  by Hermes et  al. (2012). Additionally,  we see
evidence of variability  at the following periods: $4310  \pm 200$ s
($28 \pm 3$ mma), $2309 \pm 60$ s ($25 \pm 3$ mma), $2400 \pm 120$ s
($21 \pm 3$ mma),  $4890 \pm 270$ s ($21 \pm 3$  mma), and $2094 \pm
50$ s  ($17 \pm  3$ mma). These  periodicities may  be independently
excited modes, but it is  possible that the shorter-period modes are
nonlinear combination frequencies of the longer modes. Additionally,
this  ELM  white  dwarf   has  an  unseen  companion;  spectroscopic
observations by Brown et al. (2012) find it to be in a 4.6 hr binary
with a minimum $0.64 M_{\odot}$ companion. If the ELM white dwarf is
synchronized (or nearly synchronized) with the orbital period, it is
possible that the  modes just short and long of the 4697.8 s mode
may be rotationally split components rather than independent modes.

Our $g$-mode stability  computations are in good agreement with
the range of periods observed in SDSS J184037.78+642312.3, as shown in
Fig.  \ref{noad}.  The small discrepancy  for the longest  periods can
easily  be  accommodated  taking  into account  the  uncertainties  in
$T_{\rm eff}$, which are known  to be severe because the uncertainties
characterizing  the atmosphere  models of  very low-mass  white dwarfs
(Steinfadt et al. 2012).

Finally, we  consider  an ELM  white dwarf model representative  of SDSS
J184037.78+642312.3.   From our  suite  of models,  we  found a  $0.17
M_{\odot}$  white dwarf  model with  $T_{\rm eff}=  9099$ K,  $\log g=
6.28$, and $\log(L/L_{\odot})=  0.0156$. The asymptotic period spacing
of $g$-modes  is $\Delta  \Pi^{\rm a}= 103.47$  s, and  the asymptotic
frequency spacing for $p$-modes is $\Delta \nu^{\rm a}= 4.87$ mHz.  In
Table  \ref{tabla2} we  show  the $g$-mode  periods, period  spacings,
rates of period change  and oscillation kinetic energies corresponding
to this model.  Note that  the longest observed period ($\approx 4700$
s) should correspond in our model  to a $\ell= 1$ $g$-mode with a very
high  radial   order,  $k  \approx  44-45$.   By   examining  the  radial
eigenfunction and density  of kinetic energy of these  modes, we found
that  most ($\approx 64 \%$)  of  their nodes  lay within  the He  core,
although a  significant number  of nodes (16)  are located  at regions
beyond  the He/H  interface. In  other words,  the observed  mode with
period  $\approx 4700$  s would  sample a  significant portion  of the
interior  of  the star,  besides  the  stellar  core.  The  $\ell=  1$
asymptotic  period spacing, of  $\approx 103$  s is  more than  two times
longer than for a normal $0.6  M_{\odot}$ ZZ Ceti star at $T_{\rm eff}
\approx  12\,000$  K.   So,  in  addition to  the  low  gravities  and
effective temperatures, the pulsational signatures of ELM white dwarfs
should  make  them  easy  to  distinguish them  from  the  well-known,
C/O-core ZZ Ceti stars.

\section{Summary and conclusions}
\label{conclusions}

We have  presented in this  paper the first  comprehensive theoretical
study of the seismic properties of low-mass, He-core white dwarfs with
masses   in    the   range   $0.17-0.46   M_{\odot}$\footnote{Detailed
  tabulations  of  the  the  stellar models,  chemical  profiles,  and
  pulsation periods of $g$- and $p$-modes for different stellar masses
  and   effective  temperatures   are  available   at  our   web  site
  http://www.fcaglp.  unlp.edu.ar/evolgroup.}.  We have employed fully
evolutionary  stellar   structures  representative  of   these  stars,
extracted  from the  sequences  of low-mass  He-core  white dwarfs
published by  Althaus et  al.  (2009).  These  models were  derived by
considering  the   evolutionary  history  of   progenitor  stars  with
supersolar   metallicities,  and   accounting  for   a  time-dependent
treatment  of the  gravitational settling  and chemical  diffusion, as
well as of  residual nuclear burning.  We have  explored the adiabatic
pulsation properties of these  models, including the expected range of
periods  and  period spacings,  the  propagation  properties and  mode
trapping of  pulsations, the regions  of period formation, as  well as
the dependence on the effective temperature and stellar mass.  For the
first  time, we  assessed the  pulsation properties  of  He-core white
dwarfs with masses in the range $0.20-0.45 M_{\odot}$.  In particular,
we  emphasize the expected  differences in  the seismic  properties of
objects with  $M_* \gtrsim  0.20 M_{\odot}$ and  the ELM  white dwarfs
($M_* \lesssim 0.20 M_{\odot}$).  The pulsation properties of ELM white
dwarfs  have  been  explored  recently  by SEA10.   Our  work  can  be
considered as complementary  to that study. We have  also explored the
role of time-dependent element  diffusion in ELM white dwarf models on
their  pulsational properties.   Finally,  we have  computed $g$-  and
$p$-mode blue edges of the  instability domain for these stars through
a  nonadiabatic stability  analysis.  In  particular, we  attempted to
determine if our evolutionary/pulsation models are able to predict the
pulsations exhibited  by SDSS J184037.78$+$642312.3 both  in the right
effective  temperature and  mass, and  also  in the  correct range  of
periodicities.

We summarize our findings below:

\begin{itemize}

\item  Extremely  low mass  (ELM)  white  dwarfs  ($M_* \lesssim  0.20
  M_{\odot}$) have a H envelope  that is much thicker than for massive
  He-core  white dwarfs  ($\approx 0.20-0.45  M_{\odot}$), due  to the
  very  different evolutionary  history of  the progenitor  stars.  In
  particular,  ELM white dwarfs  did not  experience diffusion-induced
  CNO flashes, thus they harbor thick H envelopes.

\item By  virtue of  the thicker H  envelope characterizing  ELM white
  dwarfs, they experience H burning via the pp chain, and consequently
  their evolution  is extremely slow.  This feature makes  these stars
  excellent candidates to  become pulsating objects (SEA10).  However,
  He-core  white dwarfs with  $M_* \approx  0.40-0.45M_{\odot}$ should
  have  evolutionary timescales of  the same  order, also  making them
  attractive targets  for current searches of  variable low-mass white
  dwarfs.

\item The  thickness of the  He/H transition region is  markedly wider
  for  the ELM  white dwarfs  than for  massive He-core  white dwarfs.
  This is due to the markedly lower surface gravity characterizing
  ELM white dwarfs, that results in a less  impact of gravitational
  settling, and eventually in a  wider chemical transition.

\item  The  Brunt-V\"ais\"al\"a  frequency  for ELM  white  dwarfs  is
  globally  lower than for  the case  of massive  objects, due  to the
  lower   gravity   characterizing   ELM   white  dwarfs.    A   lower
  Brunt-V\"ais\"al\"a  frequency  profile  leads to  longer  pulsation
  periods.

\item The bump  of $N^2$ at the He/H  transition region is notoriously
  more narrow and pronounced for the massive white dwarfs than for the
  ELM white dwarfs. This feature  results in a weaker mode trapping in
  ELM white dwarfs, something that severely limits their seismological
  potential to constrain the thickness of the H envelope.

\item As already noted  by SEA10, the Brunt-V\"ais\"al\"a frequency of
  the ELM  white dwarfs is  larger in the  core than in  the envelope.
  This is  in contrast with  the case of  models with $M  \gtrsim 0.20
  M_{\odot}$,  in  which  the Brunt-V\"ais\"al\"a  frequency  exhibits
  larger  values at the  outer layers,  thus resembling  the situation
  encountered  in ZZ  Ceti stars.  So, $g$-modes  in ELM  white dwarfs
  probe  mainly the stellar  core and  have an  enormous asteroseismic
  potential, as it was first recognized by SEA10.

\item Similarly  to ZZ  Ceti stars, the  $g$-mode asymptotic period  spacing in
  low-mass He-core white dwarfs  is sensitive primarily to the stellar
  mass,   and  to   a   somewhat  less   extent,   to  the   effective
  temperature. Specifically, $\Delta \Pi_{\ell}^{\rm a}$ is longer for
  lower  $M_*$ and  $T_{\rm  eff}$. Also,  there  is a  non-negligible
  dependence  with the  thickness  of the  H  envelope, where  $\Delta
  \Pi_{\ell}^{\rm a}$ is longer for thinner H envelopes (small $M_{\rm
    H}$). Typically, the asymptotic period spacing range from $\approx
  55$ s for $M_*= 0.45 M_{\odot}$  and $T_{\rm eff}= 11\,500$ K, up to
  $\approx 110$ s for $M_*=  0.17 M_{\odot}$ and $T_{\rm eff}= 8000$ K
  (see  Fig.  \ref{asint}).  The  evolution of  the asymptotic  period
  spacing in terms of the effective temperature is markedly influenced
  by the evolving chemical profiles due to element diffusion.

\item Similar to what happens  with the asymptotic period spacing, the
  spectrum of  $g$-mode pulsation periods themselves  is very sensitive  to the
  stellar mass, the  effective temperature and the thickness  of the H
  envelope. Again, the largest dependence is with the stellar mass.

\item  Time-dependent element  diffusion does  appreciably  affect the
  $g$-mode pulsation spectrum of  low-mass white dwarfs, in particular
  ELM  white dwarfs.  In this  regard, it  is expected  that diffusion
  processes substantially  modify the He/H chemical  interface, and so
  the resulting  period spectrum, for models with  stellar masses $M_*
  \lesssim 0.22 M_{\odot}$.   We emphasize that time-dependent element
  diffusion  should  be  taken  into  account  in  future  pulsational
  analysis of these stars.

\item  We obtained  for the  first  time the $g$-mode blue (hot)  edge of  the
  instability  domain for  He-core  low-mass white  dwarfs with  $0.20
  \lesssim M_*/M_{\odot}  \lesssim 0.45$. On the other  hand, our $g$-mode 
  blue
  edge for ELM white dwarfs  ($M_* \lesssim 0.20 M_{\odot}$) is hotter
  than that found by SEA10.  

\item We  also found a $p$-mode  blue edge of  instability of low-mass
  He-core white dwarfs. So, according to our analysis, several known
  low-mass  white dwarfs  with masses  below $\approx  0.30 M_{\odot}$
  should be pulsating stars  showing short and long periods associated
  to $p$- and $g$-modes, respectively (see Fig. \ref{tracks-zoom}).

\item Our  stability analysis successfully predicts  the pulsations 
  oberved in the  only known  variable   low-mass    white    dwarf,   SDSS
  J184037.78+642312.3  (Hermes et  al. 2012),  at the  right effective
  temperature, stellar mass and range of pulsation periods. Our 
  computations also predict the presence of short-period pulsations 
  in this star that, however, have not been detected by now.
 
\item  We found  a representative  model of  SDSS J184037.78+642312.3,
  with parameters $M_*= 0.17  M_{\odot}$, $T_{\rm eff}= 9099$ K, $\log
  g=  6.28$,  $\log(L/L_{\odot})= 0.0156$,  $\Delta \Pi^{\rm  a}=
  103.47$  s and $\Delta  \nu^{\rm a}= 4.87$ mHz. According to  
  this model,  the longest  pulsation period
  exhibited by this  star ($\approx 4700$ s) would  be associated to a
  $\ell= 1$ $g$-mode with a very high radial order, $k= 44-45$, which
  probably samples a significant portion of the star, apart from the 
  core.
\end{itemize}

The recent discovery of  SDSS J184037.78+642312.3, the first pulsating
low-mass white dwarf  star, has opened a new  opportunity to sound the
interiors of these kind of stars through asteroseismology. In the next
years a significant  number of these pulsating objects  will likely be
uncovered  through  systematic  photometric  searches  like  the  ones
performed  by Steinfadt  et al.   (2012)  and Hermes  et al.   (2012).
Needless  to  say,  in  order  to accurately  decode  the  information
embedded  in  the  pulsation  spectrum  of these  stars,  it  will  be
necessary  to  have  at hand  a  large  suite  of detailed  
evolutionary/pulsational models of low-mass white  dwarfs.  
This paper is aimed at fulfilling this requirement.

\begin{acknowledgements}
Part of this work was supported by AGENCIA through the Programa
de Modernizaci\'on Tecnol\'ogica BID 1728/OC-AR, and by the PIP 112-
200801-00940 grant from CONICET. This research has made use of 
NASA’s Astrophysics Data System.
\end{acknowledgements}

\end{document}